\documentclass[quantumrep,article,accept,moreauthors,pdftex]{Definitions/mdpi} 
\usepackage{graphicx}
\usepackage{amsthm}
\usepackage{amsmath,amssymb,amsfonts,braket}
\usepackage{comment}
\usepackage{color}
\usepackage[makeroom]{cancel}
\usepackage{marginnote}

\newcommand{\lb}{\left}
\newcommand{\rb}{\right}
\newcommand{\bs}{\begin{split}}
\newcommand{\es}{\end{split}}

\newcommand{\om}{\omega}
\newcommand{\km}{k^\mu}
\newcommand{\kom}{k^{\mu}_{(0)}}
\newcommand{\kim}{k^{\mu}_{(i)}}

\newcommand{\xma}{x^\mu}
\newcommand{\xom}{x_{(0)\mu}}
\newcommand{\xoma}{x^\mu_{(0)}}
\newcommand{\xlm}{x_{(1)\mu}}
\newcommand{\xlma}{x^\mu_{(1)}}
\newcommand{\xsm}{x_{(2)\mu}}
\newcommand{\xsma}{x^\mu_{(2)}}
\newcommand{\xim}{x_{(i)\mu}}
\newcommand{\xima}{x^\mu_{(i)}}

\newcommand{\xxi}{x_{(i)}}

\newcommand{\tti}{t_{(i)}}

\newcommand{\kx}{k_x}

\newcommand{\kxo}{k_{(0)}}
\newcommand{\omo}{\om_{(0)}}

\newcommand{\ommn}{\om_{mn}}
\newcommand{\omdr}{\om_{d}}
\newcommand{\cn}{\tilde{c}}
\newcommand{\kxi}{k_{x(i)}}
\newcommand{\omi}{\om_{(i)}}

\newcommand{\scrT}{\mathcal{F}}

\newcommand{\ampTransmittance}{a_T} 

\newtheorem{myPos}{Postulate}
\firstpage{1} 
\pubvolume{1}
\issuenum{1}
\articlenumber{0}
\pubyear{2021}
\copyrightyear{2020}
\datereceived{} 
\dateaccepted{} 
\datepublished{} 
\hreflink{https://doi.org/10.3390/quantum3010002} 

\Title{Spacetime Paths as a Whole}

\TitleCitation{Spacetime Paths as a Whole}

\Author{Sky Nelson-Isaacs \orcidA{}}

\AuthorNames{Sky Nelson-Isaacs}
\AuthorCitation{Nelson-Isaacs, S.}
\address[1]{%
Theiss Research; P.O. Box 127, La Jolla, CA 92038; theskyband@gmail.com}

\abstract{The mathematical similarities between non-relativistic wavefunction propagation in quantum mechanics and image propagation in scalar diffraction theory are used to develop a novel understanding of time and paths through spacetime as a whole. It is well known that Feynman's original derivation of the path integral formulation of non-relativistic quantum mechanics uses time-slicing to calculate amplitudes as sums over all possible paths through space, but along a definite curve through time. Here, a 3+1D spacetime wave distribution and its 4-momentum dual are formally developed which have no external time parameter and therefore cannot change or evolve in the usual sense. Time is thus seen ``from the outside''. A given 3+1D momentum representation of a system encodes complete dynamical information, describing the system's spacetime behavior as a whole. A comparison is made to the mathematics of holograms, and properties of motion for simple systems are derived.
}

\begin{document}

\section{Introduction}
In this proposal, the well-developed connection between image propagation in  scalar diffraction theory (SDT) and non-relativistic quantum wavefunction propagation (QWP) will be used to develop a 3+1D formulation of QWP and interpret the result. (Throughout this paper, 3+1D or ``four-dimensional'' refers to the usual three spatial dimensions and one temporal dimension of spacetime, or the three wavenumber dimensions ($k_x$, $k_y$, $k_z$) and one angular frequency dimension ($\om$) of momentum space.)

In Feynman's path integral formulation  \cite{FEYNMAN1948}, a spacetime path is defined as ``a sequence of configurations for successive times''. Through time-slicing, one allows the spatial path to vary over all possible spatial trajectories, but always along a specific temporal trajectory. 
The wave function $\psi(x(t))$ depends on space and time differently, where the position variable itself is assumed to be a function of time. 


The existing theory exhibits at least three shortcomings related to time that justify the need for new approaches. Firstly, the existing theory utilizes the frequency domain with great success in solving problems in the spatial domain, such as in materials science, but does not typically extend this reasoning to the temporal domain. Why not? Practically speaking, we have a very hard time thinking in a ``time-free'' manner. Are there new methods that can emerge when our technology reflects a greater equity between time and space? 
Secondly, a~related point is that time is generally understood as continuous, at least partly because we experience it that way as human subjects. The continuity of time has led to complications in physics, such as the higher order corrections to solutions of the time-dependent Schr\"{o}dinger equation, and path integrals in quantum field theory. The question arises as to whether we are thinking about time in the most expeditious way.
Thirdly, a less considered but still interesting dilemma is that the existing theory does not, in the author's opinion, adequately interpret the null interval of light. Lewis~\cite{LEWIS1926-1} pointed out a notion of ``virtual contact’’: while~starlight reaching us from far away is very old in our frame of reference, the concepts of separate emission and absorption events and ``travel’’ itself are not well defined along a null interval. What does this say about continuous time?

Here, the step is taken to examine the problem from a signal processing perspective and propose a 3+1D ``wave distribution'', ensuring symmetric treatment of space and time. 
A 3+1D wave distribution is a landscape over space and time, and thus the shape of this landscape cannot change with time because the time parameter is internal to it. 
Furthermore, a 3+1D spacetime wave distribution should also have an associated 3+1D frequency domain which has no time parameters, internal or external. Nonetheless, the~equation of motion of a system will be encoded \textit{as a whole} into the phase profile of the 3+1D wave distribution in the frequency domain. 

This leads to a novel description of a spacetime ``path as a whole''.
What is meant by a spacetime ``path as a whole''? It will be shown that a single distribution in $\km$-space corresponds to an entire path through $\xma$-space. A ``path as a whole'' is therefore a specification of a path without a specific spacetime coordinate, $\xoma$. A system's spacetime coordinates cannot be known directly from the 3+1D distribution; an interaction with a detector is required to specify the position along the path. Thus, the path predicts the spacetime coordinates at which one \textit{could} find the system, but until an interaction occurs, all spacetime coordinates along the path are equivalent.

This suggested approach addresses the previously stated shortcomings by relating the advancement of time with discrete instances of convolution. Each such convolution performs a single spacetime slice, but the slices are not required to be infinitesimally spaced, nor evenly spaced at all. Spacetime slicing happens whenever unitary evolution changes, so a single spacetime slice is synonymous with an interaction. This approach offers a ``time-free'' description of wavefunction evolution and is no different from familiar scenarios such as optical modes in a cavity.

It is well known that scalar diffraction theory (SDT) and Schr\"{o}dinger's quantum wavefunction propagation (QWP) are isomorphic to each other~\cite{FEIT1982,FLECK1977}.
The Fresnel approximation in Fourier optics and the Schr\"{o}dinger equation in quantum mechanics are both wave equations using the small angle or paraxial approximation.
The distance of longitudinal propagation of a wavefront in SDT plays a role similar to that played by time in standard QWP. 
This isometry motivates an ``image formation'' model of quantum measurement based on recursive 3+1D Fourier transforms, in which measurable time advances in discrete steps of arbitrary duration as a result of interactions. 

Quantum wavefunction collapse is thus required to be a relational or observer-dependent process. This view of measurement was part of the original interpretation of QM as early as Everett and Wigner, and, in recent years, relational collapse has become an increasingly testable hypothesis~\cite{FRAUCHIGER2018,PROIETTI2019,BONG2019}.

There are four points to be covered by the formalism. Firstly, the postulates are stated to establish the 3+1D structure based on SDT. Secondly, an operational distinction is developed between measurable "coordinate intervals" and the "background parameters" of the dual spaces. Thirdly, Feynman's path integral is written in terms of the new formalism using the isomorphism between SDT and QWP. Finally, a description of dynamics is arrived at without continual time-slicing, in favor of a novel view of ``time'' which advances discretely between interactions and spacetime paths which must be described ``as a whole''.

\subsection*{Background}

There exists an extensive catalog of research on models of time, including many that involve time as a whole. 
The two-state formalism approaches time in this way by considering both the initial and final endpoints of a system~\cite{AHARONOV1964,AHARONOV2009,COHEN2017,COHEN2018,AHARONOV2015}. In the two-state formalism, a complete description of a system is not just a single-state vector but a state vector at one time and a dual-state vector at a later time. The transition amplitude between these states, as well as the probability of states at intermediate times, depends on both the earlier and later state vectors. Events are thus contextual in time, relying on both past and future all at once.

Such ``timeless'', non-local, or as-a-whole descriptions exist in Lagrangian dynamics, quantum field theory, and Fourier optics. 
Wharton provides an as-a-whole Lagrangian method of history determination~\cite{WHARTON2014}. 
The use of light-cone gauge in string theory~\cite{ZWIEBACH2004} and quantum field theory provides a timeless and spaceless approach to simplifying certain calculations. Lewis \cite{LEWIS1926-1,LEWIS1926-2,LEWIS1926-3} argued that because photons travel along a trajectory whose Lorentz interval is identically zero, any objects exchanging photons (connected by a lightlike interval) should be considered in ``virtual contact''.
The calculation of transition amplitudes in Feynman's path integral formulation uses action functionals which are computed as a whole to determine the amplitude of an entire path. 

Vaccaro has addressed the issue of the arrow of time by proposing a unique Hamiltonian in the forward and reverse time directions, resulting in an ordered state space that represents the system at every possible time~\cite{VACCARO2016,VACCARO2018}. The Hilbert space of Vaccaro's states is, in a sense, ``timeless'', in that it describes entire histories of states which exist within an unbounded period of time. A similar work is undertaken by the proponents of the consistent histories formalism~\cite{GRIFFITHS2002}. 

One of the challenges facing both QM and QFT is the non-existence of a formal time operator which is conjugate to the Hamiltonian. Moyer has formally constructed a time operator  whose eigenstates form a basis of entire timelines (see Appendix \ref{hdr:AppTime} for a discussion) \cite{MOYER2015}. 
A formal approach to eschew externally imposed time and treat time as an internal property similar to position or momentum was developed by Pegg~\cite{PEGG1991}.
One sees spaceless calculations in SDT as well, in the use of the ``space invariant (frequency domain) amplitude impulse response'' for wavefront propagation~\cite{GOODMAN2004}. 

The parallels drawn here between SDT and QWP are well known and date from the initial development of Schr\"{o}dinger's equation from the Helmholtz equation. Joas and Lehner provide a satisfying treatment of this historical development~\cite{JOAS2009}. Schr\"{o}dinger was influenced by the optical--mechanical analogy of Hamilton, as well as Einstein's ideal gas equation of state derived from Bose--Einstein statistics, which led him to be receptive to deBroglie's proposal of the wave nature of all matter~\cite{DEBROGLIE1925}.

The formulation of quantum mechanics in terms of the Fourier transform also refers back to David Bohm's work on the ``implicate order''~\cite{BOHM1980}, and Bohm and Hiley~\cite{BOHM1993}, as well as the pilot wave formulation of quantum theory~\cite{BOHM1952-1,BOHM1952-2}. Later, de Gosson and Hiley~\cite{DEGOSSON2013} analyzed Feynman propagators in the presence of the Bohmian ``quantum potential''. This~approach uses phase space to describe as-a-whole trajectories in spacetime, reminiscent of techniques in the present proposal. Hiley~\cite{HILEY2014} examined non-commutative Clifford algebras as a foundation for the implicate order, in which the Fourier transform plays a central role.

The equivalence between scalar diffraction theory and the 3D momentum distribution of physical systems in quantum mechanics has been previously observed by Arsenault and Garcia-Martinez without addressing the time--energy relationship~\cite{ARSENAULT2001,ARSENAULT2011}.
Diffraction of wavefunctions in the temporal domain was identified by Moshinsky~\cite{MOSHINSKY1952}, and refraction of wavefunctions in spacetime was discussed more recently by J\"{a}\"{a}skel\"{a}inen, Lombard, and Zülicke~\cite{MARKKU2011}.

The paper is organized as follows. In Section \ref{hdr:formalism}, the well-known isometry between QWP and SDT is presented and applied to a novel 3+1D wave distribution in configuration and momentum space, and the main results are derived. In Section \ref{hdr:analysis}, three corollaries are derived from the postulates and implications are examined. Section \ref{hdr:closing} comments on potential further research directions and makes some comparison to the standard formalism of QM. Appendices are included with relevant background information and technical discussions.

\section{Proposed Formalism\label{hdr:formalism}}

The postulates presented here accomplish the first of our stated points: to build on the well-known isometry between diffractive image propagation and wavefunction propagation (presented in Appendix \ref{hdr:equivalence}) and formulate the evolution of 3+1D wave distributions in terms of signal convolutions. 

\subsection{Postulates\label{hdr:postulates}}

While in the QM path integral and in Hamilton's problem of minimizing classical action, time is treated as an external parameter upon which the coordinates $q(t)$ depend, in the field of signal processing, temporal data are processed in the same manner as spatial data. The two postulates presented here frame QWP using the tools of signal processing. 

\begin{myPos}
    \label{pos:p1}
The state of a system is completely characterized by wave distributions $\Psi$ and $\tilde{\Psi}$ in complex-valued dual 3+1D spaces parameterized by $\xma$ and $\km$, respectively.
\end{myPos}

This is the equivalent of a QM wavefunction. By ``system'', it is meant some signal which represents an unspecified physical entity---for instance, a Gaussian wavepacket.
Next, the convolution theorem can be expressed in two ways, between which the dual spaces are~swapped:
\begin{align}
    \label{eqn:genericP2}
    \Psi'(\xma) &= \Psi_2 \ast \Psi_1 = \scrT^{-1} \Big\{ \scrT\{\Psi_2(\xma)\} \scrT\{\Psi_1(\xma)\} \Big\} \\
    \tilde{\Psi}'(\km) &= \tilde{\Psi}_2 \ast \tilde{\Psi}_1 = \scrT \Big\{ \scrT^{-1}\{\tilde{\Psi}_2(\km)\} \scrT^{-1}\{\tilde{\Psi}_1(\km)\} \Big\},
\end{align}
where $\Psi_i$ is a given 3+1D signal, ``tilde'' indicates the distribution in $\km$-space, and $\scrT$ is the Fourier transform (or, in the general case, another appropriately chosen integral transform). 
\begin{myPos}
    \label{pos:p2}
    Together, the dual spaces comprise a complete description of the dynamical interactions between systems, obtained through convolution in either of the spaces.
\end{myPos}
For instance, in $\xma$-space (Note that the second line in Equation ($\ref{eqn:p2Xspace}$) is a special case when the distributions $\tilde{h}$ and $\ampTransmittance$ are unitary and can therefore be written as complex exponential distributions.)
\begin{align}
    \bs
    \label{eqn:p2Xspace}
    \Psi' &= h \ast (\ampTransmittance \Psi) \\
    &= \scrT^{-1} \big\{e^{-i S_k}        \scrT\big\{e^{i S_x} \Psi\big\} \big\}.
    \es
\end{align}

The sign conventions of $S_k = S_k(\km)$ and $S_x=S_x(\xma)$ are chosen for convenience, $h$ is an impulse response signal responsible for propagation through a medium, $\ampTransmittance(\xma)=e^{i S_x}$ is a spatially dependent signal known as the amplitude transmittance function or aperture function (see Goodman~\cite{GOODMAN2004}), and $\Psi=\Psi(\xma)$ is the original signal.

The central insight is that a 3+1D distribution $\Psi$ cannot evolve with respect to time and thus the requirement that $\Psi'$ equals $\Psi$.

This accomplishes the first of our goals, to state a formalization of QWP in terms of SDT and convolutions. Now, we will undertake our second goal, to distinguish between ``parameters'' and ``coordinate intervals''.

\subsection{Distinguishing \textit{Parameters} from \textit{Coordinate Intervals} \label{hdr:paramsVersusCoords}}


In 3+1D distributions, one cannot define a specific ``now'' or ``privileged present'' in $\xma$-space. This is true in special relativity as well, but for a different reason. Here, we~present a novel argument against the ``now'' which leads us to distinguish between two sets of symbols--parameters and coordinate intervals--for describing spacetime and the 3+1D frequency domain.

Consider the conversion of a sound into the frequency domain. 
Through application of the Fourier transform, one removes the explicit time dependence of a sound signal. Therefore, a sound file represented in the frequency domain cannot be time-sliced, for a given location in that file corresponds to a given frequency but has no association with a specific time. A slice therefore does not give a recognizable description of any part of the original sound. However, it is well known that the time sequencing information is still present, encoded into the phase profile of the signal, and that it can be extracted through an inverse Fourier transform. 

Similarly, $\km$-space is not parameterized by time, since the Fourier transform integrates out the time dependency. A given $\km$-space distribution therefore cannot be associated  with a particular moment in time. (Granted, one can define a 3D hypersurface in $\vec{k}$-space at a specific time using a $\delta$-function in time, for instance 
\begin{align}
    \int dx dt f(x,t) \delta(t-\tau) \exp{(-ik x+i\om t)}=\tilde{f}(k, \tau)\exp{(i \om \tau)},
\end{align}
but the resulting $\km$-space distribution has been transformed into a usual 3D wavefunction whose time dependence is the usual complex exponential dependent on a coordinate interval $\tau$ rather than a parameter $t$. To remain in the 3+1D formulation, one must avoid this ``time-slice'' into hypersurfaces. This comes at the expense of losing a meaningful notion of ``present moment'' and with the benefit of retaining information that may lead to new physics.)

Thus, on the one hand, in X-ray crystallography or in magnetic resonance imaging, the~``reciprocal lattice'' in 3D $\vec{k}$-space is used to describe the system, and a specific 3D $\vec{k}$-space distribution can be associated with each moment in time because no transform has been performed in the time domain.

However, on the other hand, what will happen to the 3+1D $\km$-space as time evolves? It~is clear that the 3+1D $\km$-space distribution cannot evolve with time because the time dependency has been integrated out.
Rather, a 3+1D $\km$-space distribution that is ``static'' can encode dynamical information about $\xma$-space within it. 
This is a restatement of Postulate \ref{pos:p2}, that $\Psi'=\Psi$.

Thus, any measurable notion of time must not be considered continuous, but rather advance in discrete steps at each moment of interaction. This is because measurement is governed by convolution, which is a discrete process applied at a particular moment.
Consider the analogous process of image propagation. To display an image involves a Fourier transform (e.g., a lens) to transform the frequency data into the spatial domain. A~setup may have multiple lenses, but never half a lens, because one cannot take half of a Fourier transform and obtain a focused image. In the same way, our notion of measurable time must advance discretely with each successive interaction, during which convolution~occurs.

As a counter-example, consider a Gaussian wavepacket with 4-momentum $\kom$, whose~$\km$-space representation is also Gaussian and is offset from the origin by ($\xxi$, $\tti$),
\begin{align}
    \bs
    \label{eqn:kGaussian1}
    \Psi &= e^{-(x-\xxi)^2/2}e^{i \kxo x - i \omo (t-\tti)} \\
    \tilde{\Psi} &= e^{-(k-\kxo)^2/2}e^{i k \xxi} \delta(\om-\omo) \lb(e^{ i \omo \tti}\rb).
    \es
\end{align}

In $\km$-space, information about motion in space is encoded into the value of $\omo$. Alternately, there could be position information encoded as a linear phase into $\tilde{\Psi}$. However, ~what is clear is that the distribution $\tilde{\Psi}$ is not changing with time, it simply encodes that the particle has energy $\omo$.
If time were to be thought of as evolving continuously in $x$-space, this~would require continually changing $\km$-space, either in phase or in amplitude. One~might imagine those changing but should keep in mind that the coordinates $\xxi$ and $\tti$ are \textit{frequencies} in that space, not parameters. Changing them would imply reference to another parameter, an external time. Instead, measurable time only exists in $k$-space when passed as a reference from $x$-space during a Fourier transform (i.e., an interaction). However,~in that case, neither space nor time can have a continual value in either space. They have values only for a given instance of the convolution operation.

Motion can be encoded within a static 3+1D $\km$-space distribution by distinguishing between the \textit{parameters} (of integration) of a space and the \textit{coordinate intervals} (or just \textit{coordinates} or \textit{intervals}) between interaction events in the space. 
Coordinate intervals are associated with interaction events, are measurable, and always have finite values. 
Parameters are associated with the Fourier integrations in Postulate \ref{pos:p2}, are unmeasurable, and always have an infinite domain.
Parameters are simply dummy variables ($x$, $t$, $k_x$, $\om$) which are necessary to convert between the dual spaces, whereas coordinate intervals (labeled with an index in parenthesis, $\xxi$, $\tti$, $\kxi$, $\omi$) are the specific measured transitions in space, time, momentum, or energy between interaction events. 
In the path integral formulation of QM, this distinction exists but it is not emphasized, as shall be shown. 

To picture the distinction between parameters and intervals, it can be useful to think about an image on a holographic film. 
A hologram is made of interference patterns capturing the phase from 2D $\vec{k}$-space onto film in 2D $\vec{x}$-space. 
What makes a hologram interesting is that these 2D interference patterns generate an image with apparent 3D coordinates. When one's vantage point on a hologram changes, the image appears to move relative to the film, i.e., the coordinates ($x_{(i)}$, $y_{(i)}$) of the image change, but the interference pattern on the film (described with parameters $x$ and $y$) does not. In a hologram, as in the theory presented here, measurable events are described by coordinates which evolve according to constraints encoded into the interference pattern described by parameters.

In the standard quantum formalism, the distinction between parameters and coordinate intervals exists but is underemphasized.
As an explicit example, consider the $n=1$ momentum--space energy eigenstate of a ``particle in a box,'' of the form
\begin{align}
    \label{eqn:particleBox}
    \tilde{\Psi}_1(k; \tau) \propto \frac{1}{1 + k L} sinc((\pi - kL)/2) e^{ik x_c} e^{-i \om_{(1)} \tau}.
\end{align}

The position parameter from the position space wavefunction has been Fourier transformed to the parameter $k$, but a time-slice has been performed in the time domain so that there are no $t$ or $\om$ parameters in the expression. Instead, they appear as \textit{coordinate intervals} in a global phase factor $\exp{(-i\om_{(1)}\tau)}$, where $\tau$ is ``measurable time''. The quantities $x_c$ and $L$ are also \textit{coordinate intervals} representing the (measurable) center of the box and its width, respectively.
Thus, the expression in Equation (\ref{eqn:particleBox}) is dependent only on the parameter $k$ and is written in $\km$-space.

Having made the distinction between what is measurable and what is unmeasurable in the formalism, we now examine how to obtain a dynamical spacetime description---an equation of motion---from within a static 3+1D $\km$-space.

\subsection{Spacetime Paths as a Whole$\label{hdr:deriveEOM}$}

Next, it will be shown that obtaining an equation of motion or dynamical mode from $\km$-space leads to a novel view of a spacetime path as a whole, and clarity will be gained on what this means.

The phases $S_x$ and $S_k$ in Equation (\ref{eqn:p2Xspace}) contain crucial dynamical information. The term ``phase map'' will be introduced here for these distributions. 
Consider example phase maps in $\km$-space (resp. $\xma$-space),
\begin{align}
    \label{eqn:simpleKPhaseMap}
    S_k(k_x,\om) &= k_x \xxi - \om \tti \\
    S_x(x,t) &= \kxi x -  \omi t,
\end{align}
where $k_x$ and $\om$ (resp. $x$ and $t$) are (unmeasurable) parameters and $\xxi$ and $\tti$ (resp. $\kxi$ and $\omi$) are coordinate intervals corresponding to an interaction. The coordinate intervals $\xxi$ and $\tti$ describe the ``frequencies'' of a plane wave in $\km$-space (often called ``spatial frequencies''). The interference patterns in the $\km$-space distribution therefore encode information about the spatial coordinate intervals of an event in $\xma$-space.
Similarly, the~interference patterns in the $\xma$-space distribution encode information about the values of 4-momentum ($\kxi$ and $\omi$) of each interaction event.

The 3+1D $\xma$-space is similar but not identical to the well-known block universe model from relativity theory. Price remarks, ``People sometimes say that the block universe is \textit{static}. This is rather misleading, however, as it suggests that there is a time frame in which the four-dimensional block universe stays the same.'' Because time is included within the block, ``it is just as wrong to call it static as it is to call it dynamic or changeable'' \cite{PRICE1996}.

The block universe presented here does not resemble physical trajectories through spacetime but instead looks like the interference patterns of a complex field which encode such trajectories.
These waves in $\xma$-space and $\km$-space are invariant with respect to their \textit{parameters}, so they do not "wave" in time. 
However, the measurable \textit{coordinate} values $\xxi$, $\tti$ of sequential interaction events \textit{can} evolve in a way which leaves the overall distributions unchanged. 
\textit{A system evolves due to discrete interactions only by coordinate intervals which leave the block universe unchanged.} These are spacetime paths as a whole.

This will now be demonstrated for a simple plane wave disturbance of momentum~$\kom$,
\begin{align}
    \label{eqn:photonXspace}
    \Psi_1(x,t) = A(x,t) e^{i \kxo x - i \omo t},
\end{align}
interacting with an ideal localizing detector, $\Psi_2(x,t)=\delta\lb(t-\tti\rb)\delta\lb(x-\xxi\rb)$, with no external potential.
Inserting these signals into Equation (\ref{eqn:genericP2}) yields
\begin{align}
    \bs
    \label{eqn:interactionXspace}
    \Psi_f(\xma) &= \scrT^{-1}\{\scrT\{\Psi_2(\xma)\} \scrT\{\Psi_1(\xma)\}\} \\
    &= \scrT^{-1}\{ e^{-i \km \xim} \tilde{\Psi}_1(\km-\kom) \} \\
    &= \Psi_1(x-\xxi, t-\tti) \\
    &= A(x-\xxi, t-\tti) e^{i(\kxo(x-\xxi)-\omo (t-\tti))},
    \es
\end{align}
where the shift property of the Fourier transform and the specific Fourier transform pair $\scrT\{\delta(x-a)\}= \exp{(-ika)}$ were used.

We derive an equation of motion by noting that \textit{a 3+1D distribution cannot evolve, thus setting $\Psi_f(\xma)=\Psi_1(\xma)$ and setting global phase factors to unity.} Setting the phase in Equations (\ref{eqn:photonXspace}) and (\ref{eqn:interactionXspace}) equal, we obtain
\begin{align}
    \label{eqn:inertialTrajConstraint2}
    \frac{\xxi}{\tti}=\frac{\omo}{\kxo}.
\end{align}

\begin{figure}[H]
    \includegraphics[width=225pt]{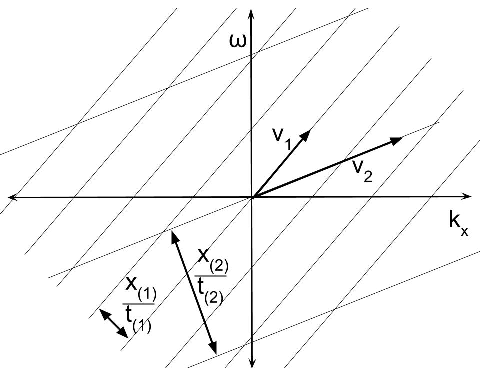}
    \caption{Discrete motion of a system. Convolution of a signal with an ideal detector is equivalent to a forward and inverse Fourier transform pair, which corresponds to a discrete update of spacetime coordinates. Shown here are two examples of such detections. For each, a phase factor applied in $\km$-space corresponds to an infinite set of consistent, discrete spacetime displacements ($\xxi$, $\tti$), associated with a well-defined velocity $v_i = \omi/\kxi$ or classical "trajectory". 
    The velocity is plotted against $\xma$-space axes (not shown).}
    \label{fig:planeWaves1}
\end{figure}

The allowable coordinate intervals ($\xxi$, $\tti$) at which a detection event will be successful are along a path of rectilinear motion with a constant velocity $\omo/\kxo$, as expected for a free system. ($\kxo$ and $\omo$ are the ``last measured'' coordinate intervals of energy and momentum.) Figure \ref{fig:planeWaves1} illustrates this geometry.

It is well-known that by applying a phase factor $e^{i\kx\xxi-i\om\tti}$ in the $\km$-space domain, a system is translated in $\xma$-space. 
Here, a novel interpretation is proposed. The translation effected by the above phase factor is a result of a forward and inverse Fourier transform pair in Equation (\ref{eqn:interactionXspace}) and is thus a discrete translation. Each such convolution effects a discrete coordinate update by a finite amount, and motion thus described is therefore not continuous but a result of discrete interaction events at arbitrary intervals.

It is now clear how a 3+1D dynamical block universe can be encoded into a static 3+1D $\km$-space.
A plane wave in $\km$-space of spatial frequency $\xima$ corresponds to a discrete transition event anywhere along a straight line path in $\xma$-space. Such a distribution in $\km$-space corresponds to an entire classical trajectory as a whole.

The equation of motion in Equation (\ref{eqn:inertialTrajConstraint2}) is a constraint on coordinate intervals, emphasizing the importance of distinguishing between parameters and coordinate intervals. This distinction also exists in Feynman's path integral formulation of QWP. We will next examine the path integral formulation to show its equivalence to the technique presented~here.

\subsection{Comparison to Feynman Path Integral Formulation \label{hdr:FeynmanPathIntegral}}

\subsubsection{Reducing to the Path Integral Formulation through Time-Slicing \label{hdr:comparePathIntegral}}

It will now be shown that the path integral formulation presented by Feynman is equivalent to applying time-slicing techniques to the formulation presented in \mbox{Postulates \ref{pos:p1} and \ref{pos:p2}.}

In Feynman's path integral, the endpoints of a path are \textit{coordinates}, whereas
complete sets of position and momentum \textit{parameter} spaces are inserted sequentially over time. Each of the position bases has an associated start and end time (coordinate interval), corresponding to a time-slice.
It is well known that time-slicing thus induces recursive forward and inverse Fourier transforms (c.f. \cite{ALBEVERIO2008}). The kernel of the transforms is derived from the Hamiltonian, factored into space- and momentum-dependent phase factors with the use of the Baker--Campbell--Hausdorff formula.

Starting from Equation (\ref{eqn:p2Xspace}), we can arrive at Feynman's time-sliced formulation by assuming a sequence of $\delta(t_j-\tau)$ convolutions arranged in a sequence for successive times. Consider the following expression for a path integral:
\begin{align}
    \label{eqn:FeynmanGeneric1}
    \bs
    \Psi(x_N, t_N) &= \int \mathcal{D} x \mathcal{D} k \,\exp{\lb(i \int_{t_i}^{t_f} (k\dot{x} - H) \, dt\rb)} \Psi(x_0,t_0) \\
    &\approx \int\int \prod_{j=0}^{N-1} d x_j \frac{d k_j}{2 \pi} \, \exp{(i (k_j\frac{x_{j+1}-x_{j}}{\tau} - H) \tau)} \Psi(x_0, t_0) \\
    &=\prod_{j=0}^{N-1}   \int \frac{d k_j}{2 \pi} \exp{(i k_j x_{j+1})} \int d x_j \exp{(-i k_j x_{j})} \exp{( -i H \tau)} \Psi(x_0, t_0)\\
    \{N=2\}&\Rightarrow \scrT_{k_1}^{-1}\{ e^{ -i \frac{k_1^2}{2m} \tau}
    \scrT_{x_1} \{ e^{ i V(x_1) \tau}
    \scrT_{k_0}^{-1}\{ e^{ -i \frac{k_0^2}{2m} \tau}
    \scrT_{x_0} \{ e^{ i V(x_0) \tau}
    \Psi(x_0, t_0) \}\}\}\}
    \es
\end{align}
where $k_j$ and $x_j$ are free parameters of momentum and position inserted at the $j$th time step, and the Hamiltonian is $H = H(k_j, x_j)$. In the second line, we have passed from an integral to a Reimann sum, where $\tau = (t_f-t_i)/N$ is a small time coordinate interval. In the last line, we have passed to an explicit representation with $N=2$, and $\scrT_x$ refers, for instance, to a transform specifically over the $x$ parameter.

It is well known that the transforms for each time-slice can be traced to the variation of Hamilton's principal function $\delta S = S( t_0+\tau)- S(t_0)$,
\begin{align}
    \label{eqn:FeynmanHamilton}
    \bs
    \Psi(x_1) &= \braket{x_1|\exp{(i\delta S)}|\Psi} \\
    &= \int\int d x_0 \frac{d k_0}{2 \pi} \, \exp{i(\frac{\partial S }{\partial x_0} \dot{x}_{01}\tau + \frac{\partial S}{\partial t_0} \tau)}
    \Psi(x_0, t_0) \\
    &= \int\int d x_0 \frac{d k_0}{2 \pi} \, \exp{\lb(i (k_0\frac{x_{1}-x_{0}}{\tau} - H(k_0,x_0)) \tau\rb)}
    \Psi(x_0, t_0)
    \es
\end{align}
where, from the theory of Legendre transforms, $k_0 = \frac{\partial S }{\partial x_0}$.

Thus, as is known, the application of phase factors $\exp{(-i k_j^2 \tau / 2 m)}$ corresponds to convolution with the non-relativistic free particle propagator, $h_x(x_j) \propto \exp{(i \frac{x_j^2 m}{2 \tau})}.$
A single time-slice of the Feynman path integral can therefore be written as 
\begin{align}
    \label{eqn:FeynmanConvolution1}
    \bs
    \Psi(x_1)  &= h_x \ast (e^{ i V(x_1) \tau} \Psi(x_0)),
    \es
\end{align}
corresponding to Equation (\ref{eqn:p2Xspace}) with $a_T = \exp{( i V(x) \tau)}$.

\subsubsection{Adding Explicit Time Domain Transforms to Path Integral Formulation \label{hdr:comparePathIntegral2}}

To apply this same analysis to the time domain, we use Equation (\ref{eqn:p2Xspace}) to write the time evolution of the wavefunction as a convolution:
\begin{align}
    \label{eqn:TimeConvolution}
    \bs
    \Psi(t_{j+1}) &= \delta(t_j - \tau) \ast \Psi(t_j) \\
    &= \scrT_\om^{-1} \{e^{i \om_j \tau} \scrT_t\{\Psi(t_j)\}\}.
    \es
\end{align}

The above uses the shift property of the Fourier transform. The updated wave distribution after one time-slice is 
\begin{align}
    \label{eqn:FeynmanFourierTransform2}
    \bs
    \Psi(x_1,t_0+\tau)&= 
    \int \frac{d k_0}{2 \pi} e^{i k_0 x_{1}} e^{ -i \frac{k_0^2}{2m} \tau}
    \int \frac{d \om_0}{2 \pi} e^{-i \om_0 t_{1}} e^{ i\om_0\tau} \\
    &\qquad \int dx_0 e^{-i k_0 x_{0}} e^{ i V(x_0) \tau}
    \int dt_0 e^{i \om_0 t_{0}} 
    \Psi(x_0, t_0)\\
    &= \scrT^{-1}\Big\{ e^{ -i \frac{k_0^2}{2m} \tau +i\om_0 \tau}
    \scrT \Big\{ e^{ i V(x_0) \tau}
    \Psi(x_0, t_0) \Big\}\Big\}.
    \es
\end{align}

The wave distribution has advanced by one time step due to the discrete shift in parameter, $t_0+\tau$. This corresponds to a single time-slice, or a single hypothetical ``time detector''. 
The phase factor written on the left in the last line can be written as the Fourier transform of an impulse response given by
\begin{align}
    \label{eqn:FeynmanImpulse}
    \bs
    \scrT \Big\{ \delta(t_0-\tau) \exp{(i\frac{x_0^2 m}{2 t_0})}\Big\} &= \int dx_0 e^{-i k_0 x_{0}}\lb(\int dt_0 e^{i \om_0 t_{0}} \delta(t_0-\tau) \exp{(i\frac{x_0^2 m}{2 t_0})} \rb) \\
    &\propto e^{ -i \frac{k_0^2}{2m} \tau +i\om_0 \tau}.
    \es
\end{align}

Therefore, a single time-slice of the non-relativistic path integral can be written as a convolution (scaling factors have been ignored):
\begin{align}
    \label{eqn:FeynmanConvolution2}
        \Psi(x_{j+1},t_{j+1}) \propto \lb( \delta(t_j-\tau) \exp{(i\frac{x_j^2 m}{2 t_j})}\rb) \ast \lb( e^{i V(x) \tau} \Psi(x_j, t_j) \rb),
\end{align}
and successive time-slices apply Equation (\ref{eqn:FeynmanConvolution2}) recursively.

Thus, the formalism in Postulate \ref{pos:p2} starts out symmetric in its treatment of time and space, but because we have enforced $N$ sequential time ``measurements'', time takes on a special role that is different from space. 
The time $\tau$ registered on each of the detectors in \mbox{Equation~(\ref{eqn:FeynmanConvolution2})} is a result of an implicit statement that a detector has interacted with the system at that time. Hence, the 3+1D Fourier transform path integral is equivalent to the Feynman path integral when convolved with a $\delta$-function in time.

These calculations suggest an alternative to time-slicing. By explicitly including both space and time domain Fourier transforms in Equation (\ref{eqn:p2Xspace}), we can derive the same results as with the time-slicing method, but the slicing is applied at finite interaction events rather than in the limit of infinitesimal slices.

\subsection{Dispersion Relations \label{hdr:dispersion}}

With the technique presented in Postulates \ref{pos:p1} and \ref{pos:p2} and the simple equation of motion of a massless free particle, Equation (\ref{eqn:inertialTrajConstraint2}), it is natural to seek dispersion relations for other familiar examples using this approach.

A wavefunction can be thought of in terms of wavepackets, or collections of frequency components of a finite bandwidth:
\begin{align}
        \label{eqn:wavepacket3D}
        \Psi(x;t) \propto \int dk \tilde{\Psi}(k) e^{-i\omega(k)t} e^{i k (x-\xxi)}.
    \end{align}
    
If time plays the role of an independent parameter upon which the spatial properties of the wavepacket depend, wavepackets will, in general, disperse because their frequency components have different speeds. 

In the 3+1D wave distribution proposed here, we note that dispersion cannot correspond to the parameters, for they are independent and unmeasurable variables of integration; just as lines on a holographic plate are fixed in space, the parameters describing the waveforms in $\xma$ and $\km$-space do not evolve in time or disperse. 

Rather, dispersion is a measurable process, so it must have to do with the coordinate intervals associated with interaction events.
We will now show how one can recover dispersion relationships for each mode of a system. We use the notion that the 3+1D state cannot change during an interaction, so any global phase that arises must vanish.

\subsubsection{Example: Non-Relativistic, Massive Free Particle}

For a non-relativistic, massive free particle, we will write the unitary evolution operator as $\hat{U}=\exp{(-i\tau \hat{p}^2/2m)} \exp{(i\tau \hat{E})}$, where $\hat{p}=-i\partial_x$, $\hat{E}=i\partial_t$ and $\tau$ is an interval of time. The time interval $\tau$ is not required to be small because the operators in the exponent commute and can be factored without the Baker--Campbell--Hausdorff relation.
Inclusion of the factor dependent on $\hat{E}$, conjugate to $t$, occurs because we are no longer treating time as the sole independent parameter.

Equation (\ref{eqn:p2Xspace}) becomes 
\begin{align}
    \label{eqn:dispersionNonRel1}
    \bs
    \Psi(x_1,t_1) &=  \braket{x_1,t_1|e^{-i\tau \frac{\hat{p}^2}{2m}} e^{i\tau \hat{E}}|\Psi} \\
    &=  \int \int dk_0  d\om_0 \braket{x_1,t_1|e^{-i\tau \frac{\hat{p}^2}{2m}}e^{i\tau \hat{E}}|k_0,\om_0} 
     \int  \int dx_0 dt_0 \braket{k_0,\om_0|x_0,t_0}  \Psi(x_0, t_0) \\
    &=  \int \int dk_0  d\om_0 
    e^{(ik_0 x_1 - i \om_0 t_1)} e^{-i\frac{k_0^2 \tau}{2m}}e^{i\tau \om_0}
    \int \int dx_0 dt_0 e^{(-ik_0 x_0 + i \om_0 t_0)}\Psi(x_0, t_0)
    \es
\end{align}

The eigenfunctions of this transformation are plane waves, so the general form of the 3+1D wave distribution is
\begin{align}
    \label{eqn:KGPlaneWaves}
    \Psi(x_0,t_0) = \sum_n c_n e^{i k_{(n)} x_0 - i \om_{(n)} t_0},
\end{align}
where $k_{(n)}$ and $\om_{(n)}$ are coordinate intervals, i.e., single-valued, measurable properties of the system.

Inserting Equation (\ref{eqn:KGPlaneWaves}) into Equation (\ref{eqn:dispersionNonRel1}),
\begin{align}
    \label{eqn:dispersionNonRel2}
    \bs
    \Psi(x_1,t_1) &=  \int \int dk_0  d\om_0 
    e^{i k_0^\mu x_{1\mu}} e^{-i\frac{k_0^2 \tau}{2m}}e^{i \om_0 \tau}
    \int \int dx_0 dt_0 e^{-i k_0^\mu x_{0\mu}} \sum_n c_n e^{i k_{(n)} x_0 - i \om_{(n)} t_0} \\ 
    &= \sum_n c_n \int \int dx_0 dt_0 \int \int dk_0 d\om_0 \, e^{i k_0^\mu  (x_{1\mu}-x_{0\mu})}e^{-i \frac{k_0^2}{2m}\tau}e^{i \om_0 \tau}
     e^{i k_{(n)} x_0 - i \om_{(n)} t_0} \\
    &= \sum_n c_n \int \int dx_0 dt_0 \braket{x_1|x_0} \braket{t_1|t_0}
    e^{-i\tau \frac{(-i\partial_{x_0})^2}{2m}} e^{i\tau (i\partial_{t_0})} e^{i k_{(n)} x_0 - i\om_{(n)} t_0} \\
    &= \sum_n c_n e^{i k_{(n)} x_1 - i\om_{(n)} t_1} e^{-i\frac{k_{(n)}^2}{2m} \tau } e^{i\om_{(n)} \tau}. \es
\end{align}

The first factor in the last line is just the original wavefunction decomposition, \mbox{Equation (\ref{eqn:KGPlaneWaves})}, so the global phase factors on the right must be set to unity for each mode $n$ in order for the wavefunction to be invariant. This is true if
\begin{align}
    \bs
    \om_{(n)} = \frac{k_{(n)}^2}{2m}.
    \es
\end{align}
This is the usual non-relativistic dispersion relationship for a mode $n$.

Next, this process will be repeated for the quantum harmonic oscillator, an exactly solvable system of central importance in quantum mechanics and quantum field theory.

\subsubsection{Example: Quantum Harmonic Oscillator}

The well-known dispersion relationship for each mode of a quantum harmonic oscillator can be reproduced using the new formalism by applying Equation (\ref{eqn:p2Xspace}) over a single quantum transition event. Equation (\ref{eqn:p2Xspace}) is
\begin{align}
    \label{eqn:HarmOscTransitionAmp}
    \Psi(x_1,t_1) = \braket{x_1,t_1|\exp{\lb(-i \hat{H}\tau\rb)}e^{i\tau \hat{E}}|\Psi},
\end{align}
where the Hamiltonian is 
\begin{align}
    \label{eqn:HarmOscHamiltonian}
    \hat{H}=\frac{\hat{p}^2}{2m} + \frac{m\Omega^2}{2} \hat{x}^2.
\end{align}

The time interval $\tau$ is not required to be small because, as will be seen, we will not need to factor non-commuting operators in the exponent.
We will assume separability in the space and time dependence of the wave distribution, $\Psi(x,t) = \psi(x) \phi(t)$. We insert complete bases of $\ket{\om}$ and $\ket{t}$ eigenstates as before, but instead of inserting complete bases of $\ket{k}$ and $\ket{x}$ eigenstates separately, we insert a complete basis of energy eigenstates, $\sum_n \ket{n}\bra{n}$. The time dependence is 
\begin{align}
    \label{eqn:timeEigenstates}
    \bs
    \phi_n(t) = e^{-i \om_{(n)} t}
    \es
\end{align}
found from the eigenfunctions of the equation 
    \begin{align}
        \bs
        \label{eqn:timeDependence}
        \phi'(t) &= \int d \om_0 \braket{t_1| e^{i \hat{E}\tau}|\om_0} \int dt_0 \braket{\om_0|t_0}\braket{t_0|\phi}  \\
        &= \int d \om_0 e^{-i \om_0 t_1} e^{i \om_0 \tau} \int d t_0 e^{i \om_0 t_0}  \phi(t_0),
        \es
    \end{align}
which has the same form as Equation (\ref{eqn:TimeConvolution}). The states $\ket{\om_0}$ are eigenstates of the operator $\hat{E}$. 

Defining $c_n \equiv \braket{n|\psi}$, $\psi_n \equiv \braket{x_1|n}$, and $\braket{t_0|\phi} \equiv \exp{(-i \om_{(n)} t_0)}$,
Equation (\ref{eqn:p2Xspace}) is then
\begin{align}
    \label{eqn:HarmOscMethodB}
    \bs
    \Psi(x_1,t_1) &= \sum_n \int d \om_0 \braket{x_1, t_1|e^{-i\hat{H}\tau}e^{i \hat{E} \tau} |n, \om_0} \int dt_0 \braket{n|\psi}\braket{\om_0|t_0}\braket{t_0|\phi} \\            
    &= \sum_n c_n \int d\om_0 \braket{x_1|e^{-i\hat{H}\tau}|n}
    \braket{t_1|e^{i \hat{E} \tau} |\om_0} \int dt_0 e^{i \om_0 t_0} e^{-i \om_{(n)} t_0} \\
    &= \sum_n c_n \braket{x_1|n} e^{-i\tau \hbar \Omega (n+\frac{1}{2})} \int d \om_0 e^{-i \om_0 t_1}e^{i \om_0 \tau} \delta(\om_0-\om_{(n)})\\
    &= \sum_n c_n \psi_n \phi_n \exp{(-i \hbar \Omega (n+1/2) \tau + i \om_{(n)} \tau)},
    \es
\end{align}
where the well-known relation $\hat{H}\ket{n} = \hbar \Omega (n+1/2)$ for the harmonic oscillator was used.

Once again, we require the wave distribution to be invariant throughout this calculation since there is no external time for it to change with respect to.
Equation (\ref{eqn:HarmOscMethodB}) shows that this occurs if the  global phase factor for each mode vanishes. This constraint reproduces the well-known energy spectrum of the harmonic oscillator,
\begin{align}
    \om_{(n)}= \hbar\Omega(n+1/2).
\end{align}

Factors of $\hbar$ have been explicitly reintroduced for clarity. 

\subsection{Example: Two State System \label{hdr:twostateSystem}}

We now explore the application of Equation (\ref{eqn:p2Xspace}) to a two-state system. We will consider the dynamics of a spin-1/2 particle driven by an oscillating potential. This problem is typically solved by applying perturbation theory to the time-dependent Schr\"{o}dinger equation (TDSE). The analysis will illustrate the equivalence of the present method to the usual perturbation theory approach, as well as provide a novel consideration when thinking about time dependence in equations of motion.

The energy eigenstates are those of the unperturbed Hamiltonian, $\hat{H}_0 \propto \hat{S}_z$, which~has two states $\ket{\pm}$ for spin along the $\hat{z}$-axis. The perturbation is characterized by a time-dependent driving potential, 
\begin{align}
    \bs
    \label{eqn:TwoStatePotential}
    V_{mn} &= V_0 e^{-i\omdr t}.
    \es
\end{align}
The aim is to find the solution to Schr\"{o}dinger's equation, $\ket{\Psi(t)} = c_1(t)\ket{+} + c_2(t)\ket{-}$.

The standard approach is to insert the time-dependent Hamiltonian 
\begin{align}
    \bs
    \label{eqn:}
    \hat{H}_0 + \hat{V}(t) \rightarrow \begin{pmatrix} E_1 & V_{12}(t) \\
     V_{21}(t) & E_2
    \end{pmatrix}
    \es
\end{align}
into Schr\"{o}dinger's equation, obtaining an equation for the $m$, $n$ energy eigenstates,
\begin{align}
    \bs
    \label{eqn:standardTwoState}
    i \hbar \dot{c}_m(t) &= \sum_n V_{mn}(t) e^{i \ommn t} c_n(t),
    \es
\end{align}
where $\ommn = \om_m-\om_n$. 
One can proceed by integrating this equation with respect to time from the initial state to the final state,
\begin{align}
    \bs
    \label{eqn:integrateSchrodinger}
    c_m(t) &= -\frac{iV_0}{\hbar} \int_0^\tau dt \sum_n e^{i \ommn t} e^{-i \omdr t} c_n(t),
    \es
\end{align}
obtaining an expression for the states' time dependencies.

We shall now show that this is equivalent to Equation (\ref{eqn:p2Xspace}) but that the new approach provides additional insight into the nature of time in the calculation.
Two adjustments are necessary to show this equivalence. Firstly, 
we will redefine the potential with a windowing function,
\begin{align}
    \bs
    \label{eqn:potentialTwoState}
    a_T \equiv V_{mn}'(t) &= V_0 e^{-i \omdr t} rect(\frac{t-\tau/2}{\tau}).
    \es
\end{align}

This is reasonable if we are thinking of signals in $\xma$- or $\km$-space, and this step allows us to extend the time integral in Equation (\ref{eqn:integrateSchrodinger}) across the entire domain without changing the value of the integral.

Secondly, Postulate \ref{pos:p1} states that the wave distributions live in $\xma$-space and $\km$-space, so it is clear that the coefficients $c_n$ cannot be parameterized by both time \textit{and} energy (see~the discussion on parameters in Section~\ref{hdr:paramsVersusCoords}). Rather, they are part of either one space \textit{or} the other space. Since they are coefficients of the energy eigenstates, which live in $\km$-space, they~must be parameterized by energy, $\om_n$. The distinction between parameters and coordinates becomes important here because parameters and coordinates evolve in different ways. The~``time dependence'' of these coefficients actually refers to a time \textit{coordinate}. This means that the coefficients are not updated continuously but \textit{iteratively} through convolution.

Accordingly, we will relabel $c_m(t) \rightarrow \cn_m(\om_m)$, the tilde emphasizing that the symbol lives in $\km$-space (i.e., $\om$-space).
With these two adjustments, the integral turns into a Fourier series converting from the $\om$ domain to the time domain and a Fourier transform converting back to the $\om$ domain. 
\begin{align}
    \bs
    \label{eqn:convolutionTwoState}
    \cn_m(\om_m) &= -\frac{iV_0}{\hbar}\int dt e^{i\om_m t} e^{-i \omdr t} rect(\frac{t-\tau/2}{\tau})\sum_n e^{-i \om_n t}  \cn_n(\om_n)
    \es
\end{align}

Here, the basis states used in the inverse transform are the two $\ket{\pm}$ eigenstates of $\hat{S}_x$, labeled by $m,n$, rather than a continuous distribution over $\om$.
This is the first example we have seen of a quantum superposition of states in this theory. It occurs because we are transforming into a basis other than $\km$ or $\xma$, and the Fourier transform is a linear operator so it allows linear superpositions. The key distinction is the transform over $\ket{\pm}$ instead of over $\ket{\om}$.

Now that we have it in this form, the remainder of the example follows the standard derivation. We will find the resonant modes of the system by evaluating the integral in the time domain first, where the symbols $\scrT$ and $\sum$ emphasize the transform operations between the dual spaces.
\begin{align}
    \bs
    \label{eqn:mainFourierTwoState}
    \cn_m(\om_m) &= \scrT\{... \Sigma_n\{...\}\} \\
    &= -\frac{iV_0}{\hbar}  \int_{-\infty}^{\infty} dt  e^{-i\omdr t} rect(\frac{t-\tau/2}{\tau})\sum_{n} e^{i \ommn t} \cn_n \\
    &= -\frac{iV_0}{\hbar} \sum_{n} \cn_n  \int_{0}^{\tau} dt e^{i \ommn t} e^{-i\omdr t} \\
    &= -\frac{iV_0}{\hbar} \sum_{n} \cn_n e^{i(\omdr-\ommn)\tau/2} \frac{\sin{\lb(\frac{(\omdr-\ommn)}{2}\tau\rb)}}{(\omdr-\ommn)/2} \\
    \es
\end{align}

Equation (\ref{eqn:mainFourierTwoState}) is equivalent to the results from first-order perturbation theory. Here, we have applied exactly one forward and inverse transform between $\xma$-space and $\km$-space. This is a matrix equation whose eigenstates can be solved for the coefficients $\cn$ via the usual methods. It suffices here to use the zeroth order values of the $\cn$ as the initial conditions to eliminate $\cn_n$, obtaining for the coefficient of $\ket{-}$,
\begin{align}
    \bs
    \label{eqn:mainFourierTwoStateFinal}
    \cn_2 &= -\frac{iV_0}{\hbar} e^{i((\omdr-\ommn)\tau/2)} \frac{\sin{\lb(\frac{(\omdr-\ommn)}{2}\tau\rb)}}{(\omdr-\ommn)/2}. \\
    \es
\end{align}

It should be noted that our result is not parameterized by time. The coordinate $\tau$ that appears is a result of the interaction with the potential/aperture in Eqn. $\ref{eqn:potentialTwoState}$. Another interaction will update this value, but it does not vary continually. One should not think of time as evolving in the expression (\ref{eqn:mainFourierTwoStateFinal}), since it is written in $\km$-space.

Equation (\ref{eqn:mainFourierTwoStateFinal}) is the same distribution in $\km$-space found by typical methods, and the resonant driving frequency $\omdr=\ommn$ is easy to read off the equation. Through a single Fourier transform of the windowed potential $a_T$, we are able to characterize the system's resonant modes.

\subsubsection*{Example: Coulomb Potential}

Here, a brief suggestion will be made on how to apply this approach to the Coulomb potential, $V(\vec{r})=\frac{1}{r}$, but the case will not be fully worked out. We use the second line of Equation (\ref{eqn:p2Xspace}), in this case using spherical coordinates. A complete set of spherical harmonic momentum eigenfunctions is inserted, 
\begin{align}
    \label{eqn:kernelFunctions}
    \braket{\textbf{k}_{nlm},\om|\textbf{r},t} = j_l(k_n r) Y_l^m(\mu,\phi) e^{-i\om t},
\end{align}
instead of Cartesian momentum eigenfunctions. Postulate \ref{pos:p2} takes the form
\begin{align}
    \label{eqn:coulombPot}
    \bs
    \Psi'(\vec{r'}) &\propto \sum_{\substack{nlm}} \braket{\vec{r'}|\hat{H}(\hat{p})|\textbf{k}_{nlm},\om} \int dr\, d\Omega\, dt \braket{\textbf{k}_{nlm},\om|e^{i\hat{V}(r)\delta t}|\textbf{r},t}\braket{\textbf{r},t|\Psi},
    \es
\end{align}
so the conversion between the dual spaces is the spherical harmonic transform instead of the Fourier transform.
The operator $\exp{(i\hat{V}(r)\delta t)}$ acting on these basis functions results in a $\xma$-space phase factor whose phase is a function of the eigenvalues of this operator, which are the quantum numbers $n$, $l$, $m$. As before, the invariance of the phase distribution constrains the coordinates and determines the equation of motion.

Interestingly, the inverse transform over the spherical harmonics will be discrete, with the basis states labeled by $n$, $l$, and $m$. Under this condition, the ``coordinates'' in $\km$-space are not ``points'' in a continuous $\km$-space but the quantum numbers $n$, $l$, and $m$ associated with the discrete distributions $j_l(k_n r) Y_l^m(\mu,\phi)$. In other words, ``locations'' in the space correspond to the various harmonics that can exist for the system. 

The simple results of the hydrogen atom seem trivial to reproduce with this approach, since they are just the spherical harmonic basis states, but the tools of signal processing suggested here may prove useful in reproducing the more complicated harmonic electronic states of heavier elements.

\section{Analysis\label{hdr:analysis}}


In Section~\ref{hdr:formalism}, it was shown that formulating quantum wavefunction propagation in the language of diffraction theory and holograms leads to a distinction between unmeasurable parameters and measurable coordinate intervals, can be used to derive equations of motion and dispersion relations of a few sample cases, and reduces to a standard form of the path integral when one applies time-slicing. Now, we will aim to interpret these results to understand what time evolution and motion mean in this formalism.

\subsection{Corollaries: Spacetime as a Whole \label{hdr:results}}

The argument that spacetime can be described as a whole will be formalized with three corollaries derived from Postulates \ref{pos:p1} and \ref{pos:p2}, relating to the reality (or lack thereof) of quantum systems and descriptions of spacetime as a whole. (These results reconstruct the thought experiment devised by Wheeler on delayed choice, but based on a unique premise.)



\begin{Corollary}
    \label{cor:c1}
    The distributions $\Psi$ and $\tilde{\Psi}$ in the respective $\xma$- and $\km$-spaces must be invariant between interactions because $\km$-space is not parameterized by time.
\end{Corollary}

Corollary \ref{cor:c1} was stated immediately after the postulates in Section \ref{hdr:postulates} and used to derive equations of motion.

\begin{Corollary}
    \label{cor:c2}
    Given a system represented by $\Psi(\xma)$, its path of travel (i.e., the set of admissible coordinates where a successful measurement could occur) is defined in $\km$-space as a whole.
\end{Corollary}

\begin{Corollary}
    \label{cor:c3}
    Only interactions are assigned physically meaningful descriptions, i.e., coordinate intervals, and no physically meaningful description of unitary evolution can be made between these. 
\end{Corollary}

It has already been argued that a 3+1D wave distribution cannot vary with time because $x$ and $t$ are both independent parameters and there is no other ``time-like'' parameter with respect to which the distributions (in particular, $\km$-space) could vary. The 3+1D $\xma$-space distribution must be an immutable landscape over space and time parameters, and similarly for $\km$-space. This is the content of Corollary \ref{cor:c1}.

To examine Corollaries \ref{cor:c2} and \ref{cor:c3}, consider the 1+1D plane wave detected after travelling across a coordinate interval $\xxi$, $\tti$ in Equation (\ref{eqn:interactionXspace}). 
The expression inside the inverse Fourier transform on the second line provides an expression for the system in $\km$-space,
\begin{align}
    \label{eqn:spacelessKspace}
    \tilde{\Psi}_{f,1}(\km) = e^{i(\kx\xxi-\om \tti)} \tilde{\Psi}_1(\kx-\kxo,\om-\omo),
\end{align}
which is explicitly dependent on the locations of the boundaries in space and time, i.e., the~interval of transit ($\xxi$, $\tti$). This is a result of the integral transform relationship between the dual spaces.
Because the distributions are invariant between interactions (Corollary \ref{cor:c1}), this particular distribution in $\km$-space corresponds to the entire span between interactions, i.e., an entire path of travel through $\xma$-space for the system. This is \mbox{Corollary \ref{cor:c2}.}

\begin{figure}[H]
  
  \includegraphics[width=225pt]{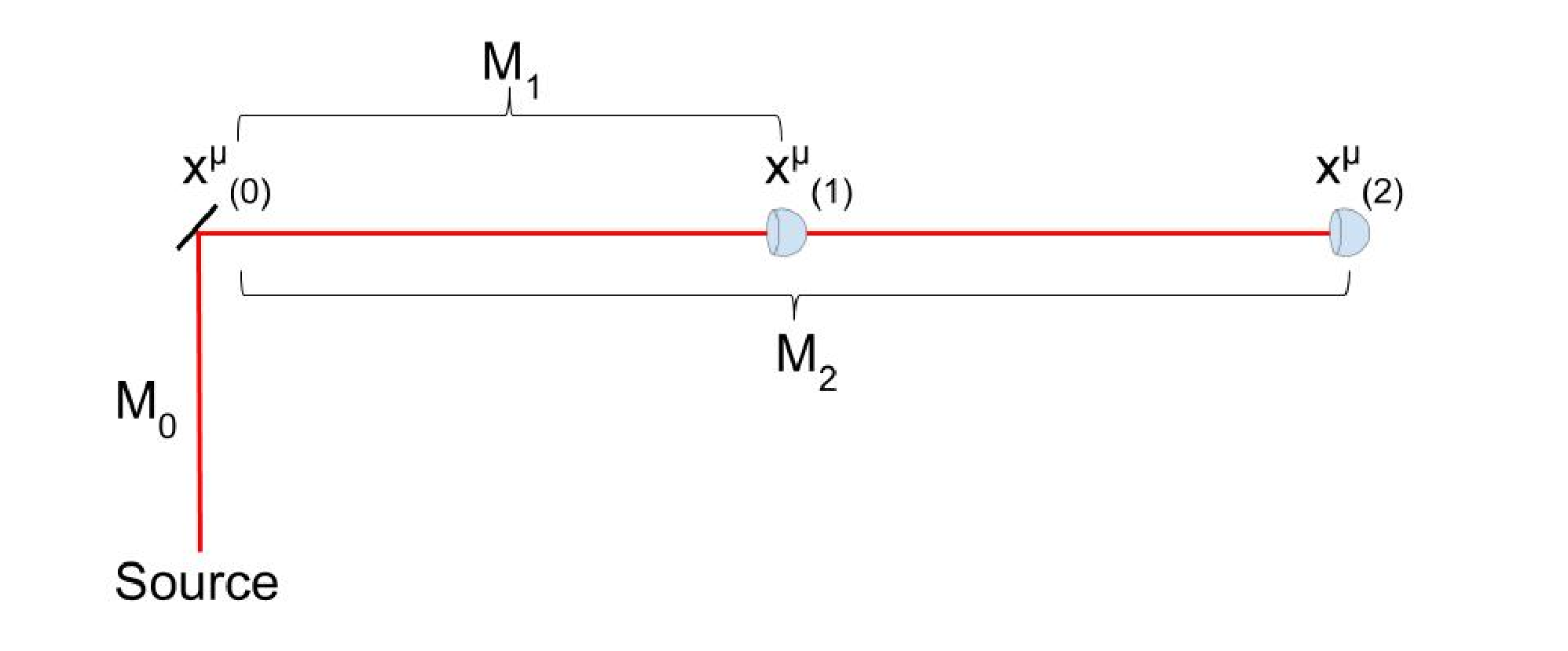}
  
  \caption{A photon travels according to a $\km$-space map $M_0$, from a source to a mirror, and is then redirected onto either a path described by map $M_1$ or a path described by map $M_2$. Each of these is a unique distribution in $\km$-space due to its unique endpoints, $\xlma$ or $\xsma$. The ``delayed choice'' between $M_1$ and $M_2$ is made by inserting or removing the detector at $\xlma$ \textit{after} the event at $\xoma$~occurs.}
  \label{fig:branches}
\end{figure}

To illustrate this further, Figure \ref{fig:branches} shows that the interval $M_1$ with endpoints $\xom$ and $\xlm$ corresponds to Equation (\ref{eqn:spacelessKspace}) with $i=1$. Should the system take a different path through $\xma$-space, a different $\km$-space distribution would be required because each $\km$-space distribution depends explicitly on the coordinate interval between start and endpoints. 
For instance, if the detector at $\xlm$ is removed, the path from $\xom$ to $\xsm$ properly describes the system, and the proper expression is Equation (\ref{eqn:spacelessKspace}) with $i=2$. 
        
$\tilde{\Psi}_{f,1}$ (resp. $\tilde{\Psi}_{f,2}$) describes a whole path of motion from $\xom$ to $\xlm$ (resp. $\xsm$).
What happens to the $\km$-space representation, $\tilde{\Psi}_{f,1}$, when the experimenter's choice occurs? Does $\tilde{\Psi}_{f,1}$ change to $\tilde{\Psi}_{f,2}$? This cannot be so, because $\km$-space cannot change in time. A~particular configuration of $\km$-space cannot be associated with a particular time, like~a snapshot of a filmstrip, so the time at which the experimenter's choice is made cannot affect whether $\tilde{\Psi}_{f,1}$ or $\tilde{\Psi}_{f,2}$ applies to the motion. Both distributions must apply; the final measured result must be determined only by the final interaction at either $\xlm$ or $\xsm$. Each of these outcomes, paired with the initial emission point, forms a distinct path as a whole. This means that the $\km$-space distribution corresponds to the entire interval and cannot evolve during it.

Corollary \ref{cor:c2} thus follows from Postulate \ref{pos:p1}, because in an integral transform relationship, specific regions of $\km$-space cannot correspond to specific regions of $\xma$-space, e.g., specific times. The system's entire $\xma$-space behavior, including beginning and endpoints of the system's trajectory, corresponds to a single distribution in $\km$-space. 

Image processing and holography again provides a useful illustration of the main point. It is well known that the frequency domain representation of a photographic image contains a complex spectrum that has integrity as a whole and cannot be subdivided. In~other words, cutting the photograph physically in half alters the entire complex spectrum. The frequency domain describes relationships across images as a whole. Isolating a subdivision of an image or signal introduces non-local artifacts into the frequency domain representation. 

This is an important fact in the consideration of 3+1D wave distributions. The trajectories or paths of these signals are determined by a given 3+1D $\km$-space distribution. Just as locally altering the complex spectrum of an image in 2D $\vec{k}$-space generates artifacts or changes to the entire image in 2D $\vec{x}$-space, a local modification of the $\km$-space representation of a signal will affect the spatial and temporal characteristics of an entire chain of events in spacetime.

Corollary \ref{cor:c3} comes from the observation that the parameters are unmeasurable. Only~the coordinate intervals are measurable, such as $\xxi$, $\tti$ in Equation (\ref{eqn:spacelessKspace}); thus, only the endpoints of a path are describable in physical terms. 
The path of a particle moving across an interval $\xim$ will have a distribution in $\km$-space given by \mbox{Equation (\ref{eqn:spacelessKspace})}, which contains no reference to $\xma$-space other than the overall interval of travel, $\xim$. This~$\km$-space expression for the particle therefore contains no information describing continuous movement of the particle in $\xma$-space, only its \textit{overall interval} of travel. The~Fourier transform is unitary, so information is preserved during the transform between the spaces. Thus, the lack of $\km$-space information about the continuous movement of the particle in space means that such information cannot exist in $\xma$-space either. All that exists is the information about the overall interval of travel. Therefore, the interactions which initiate or terminate a path of travel have a physical description, while the unitary evolution describing the path of travel does not have a physical description.

Thus, the paths of travel are defined as a whole and dynamics can be formulated without time-slicing. The motion of a system through spacetime is composed of discrete leaps between interactions rather than continuous evolution in time.

\subsection{Consistency with Known Experiment \label{hdr:falsifiability}}

The postulates and corollaries require that interaction is observer-dependent. By observer dependence, it is meant that each measurement generates correlations between the observing system and the observed system but leaves the combined system in a superposition state from the vantage point of an outside observer. The proposal here appears to be falsifiable in this regard and can in principle be experimentally distinguished from some other contending formulations of quantum mechanics. 

Distinguishing observer-dependent from observer-independent measurement has recently undergone advances in testability. Any experiments which support observer dependence will be consistent with this formalism. 

How does the model predict observer-dependent state collapse? 
It has been argued and concluded in Corollary \ref{cor:c1} that $\km$-space cannot change (in the absence of a separate, auxiliary time parameter). However, it is also clear from the vantage point of any individual observer that systems do change. Such changes must be relative to the observer. Then, the 3+1D wave distribution will not change relative to its parameters, even while interactions within the environment occur, tracked by coordinate intervals.

Thus, change can happen from an observer-dependent perspective, but such change must be described relationally, while the wave distribution globally does not change.

The Wigner's Friend thought experiment explores observer-dependent interactions. In this setup~\cite{WIGNER1961}, an observer $O1$ measures a system $S$ and a second observer $O2$ measures system $S+O1$. If the second observer $O2$ obtains statistics for the system that describe a mixed state, then collapse must have ``really happened'' (observer-independent). If, instead, statistics are obtained that describe interference, then the expanded system $S+O1$ is in a joint superposition (observer-dependent).

Recent experimental proposals~\cite{FRAUCHIGER2018,BRUKNER2018} and results~\cite{PROIETTI2019,BONG2019} are consistent with observer-dependent collapse and are therefore consistent with the proposed formalism. Two of these will be discussed.

Proietti et al. used six entangled photons to track two entangled systems each consisting of ``photon'' (System), ``observer'' (Friend), and ``super-observer'' (Alice/Bob).
Their results are consistent with ``no-collapse'' quantum evolution: the System and the Friend remain in a superposition state even with confirmation of Alice's Friend's definite measurement.

According to their analysis, the results indicate that at least one of the three following assumptions must be false:

\begin{itemize}
    \item (L) locality;
    \item (F) free choice;
    \item (R) observer-independent facts.
\end{itemize}

Observer-dependent collapse occurs when we retain the first two assumptions above while discarding the third. 
If one feels confident in locality and free choice, then this experiment requires collapse to be a relative process, consistent with the formalism presented here.

A related paper by Frauchiger and Renner \cite{FRAUCHIGER2018} proposes an experiment which shows that QM generates inconsistent experimental predictions if, again, three assumptions are all held to be true: 

\begin{itemize}
    \item (Q) universal application of quantum probabilities (without a dividing line between classical and quantum domains);
    \item (C) consistency of measurements between observers;
    \item (S) objectivity of the measurement and collapse process.
\end{itemize}

Although the details differ, in Frauchiger and Renner's analysis, if one feels confident that quantum mechanics is a universal theory and that consistency of measurement results between observers is valid, then the thought experiment implies violation of ``observer-independent collapse'' and is consistent with the theory presented here.

\section{Discussion \label{hdr:closing}}

Further avenues of inquiry are suggested by this approach.

One might wonder whether this theory can accommodate multiple particles. The~approach does not formally define particles, for the $\xma$- and $\km$-space distributions do not correspond to individual entities. This can be seen in that the convolution in \mbox{Equation (\ref{eqn:p2Xspace})} takes two signals as input and returns one signal as output, indicating that individual systems are not fundamentally distinguishable.
It might be possible to formally connect the $\xma$- and $\km$-space distributions with the field operators $\hat{\phi}$ and $\hat{\Pi}$ of quantum field theory, with the usual creation and annihilation operators generating multi-particle disturbances in the distributions.

\subsection{Noether's Theorem and Wavefunction Collapse \label{hdr:appNoether}}

When one applies Noether's theorem to space and time translation in $\xma$-space, one obtains the laws of momentum and energy conservation; however, when Noether's theorem is applied to boost invariance in $\km$-space, the result seems less impressive, just the equation of motion of the center-of-mass. Here, it is argued that Noether's theorem applied to symmetries under shifts in $\km$ coordinates results in a ``conservation of trajectory'' and can be associated to (relational) quantum collapse.
Thus, collapse of a linear system to one of its eigenvectors can be seen as a conservation law.

Noether's theorem provides a link between physical symmetries and conservation laws. A conservation law will exist for a property if the Lagrangian $\mathcal{L}$ is not dependent on that property, i.e., if $\mathcal{L}$ is not a function of time, then energy is conserved over time, and similarly for the space/momentum pair. If both are true, then Noether's theorem results in the relation ~\cite{NEUENSCHWANDER2011}
\begin{align}
    \label{eqn:NoethersTheorem}
    p_\mu \xi^\mu - H\tau = \text{constant},
\end{align}
where $\xi^\mu$ and $\tau$ are generators of translation. Symmetries in $\xma$-space result in conserved coordinates in $\km$-space. What does this look like in $\km$-space? Just as invariance of the action under infinitesimal translations in $\xma$-space leads to conserved quantities in $\km$-space, invariance of the phase distribution in $\km$-space under infinitesimal boosts leads to a conservation law in $\xma$-space. The result thus obtained is known as the center-of-mass theorem.  As in Equation (\ref{eqn:NoethersTheorem}), space and time are not individually conserved, but a certain relationship between them is indeed conserved, a ``conservation of trajectory''.

We emphasize that symmetries of motion involve coordinates (which are measurable), not parameters.
We examine the phase map in $\km$-space as we vary the coordinates $\kim$. Through the usual methods, we expand the derivative in terms of its dependencies on the coordinates in $1+1D$ $\km$-space,
\begin{align}
    \bs
    \delta S_k(\kim) &= \frac{\partial S_k(\kim)}{\partial \kxi} \delta \kxi + \frac{\partial S_k(\kim)}{\partial \omi} \delta \omi\\
    &= 0.
    \es
\end{align}
$\frac{\partial S_k}{\partial \kim}=\xim$ is the conjugate of $\kim$, a specific $\xma$-space coordinate associated with the distribution.

As an explicit example, consider a Gaussian wavepacket with momentum $\kxi$, which strikes a half-silvered mirror and partially reflects directly backward, picking up a component with momentum $-\kxi$. Its distribution in $\km$-space consists of two terms of the form
\begin{align}
    \bs
    \label{eqn:NoetherGaussian1}
    \frac{1}{\sqrt{2}}e^{-(\kx\pm \kxi)^2/2}e^{i (\kx \pm \kxi) \xxi - i \omi \tti}
    \delta(\om-\omi) 
    \es
\end{align}

The distribution now displays two pulses, with positive and negative values of $\kxi$, respectively.
Upon varying the coordinates $\kxi$ and $\omi$, the phase distributions of the two pulses vary. 
Given phase profiles
\begin{align}
    \bs
    S_1 &= (\kx+\kxi) \xxi - \omi \tti\\
    S_2 &= (\kx-\kxi) \xxi - \omi \tti,
    \es
\end{align}
the variations are
\begin{align}
    \bs
    \label{eqn:variationS1}
    S_1(\kxi + \delta \kxi, \omi + \delta \omi) &= S_1(\kxi, \omi) + \delta S_1( \delta\kxi, \delta\omi) \\
    &= S_1(\kxi, \omi) + \delta \kxi \xxi - \delta\omi \tti
    \es
\end{align}
and 
\begin{align}
    \bs
    \label{eqn:variationS2}
    S_2(\kxi + \delta \kxi, \omi + \delta \omi) &= S_2(\kxi, \omi) - \delta \kxi \xxi - \delta\omi \tti,
    \es
\end{align}
where the difference is a minus sign in front of $\delta \kxi$.

The variation can vanish for the first pulse if $\delta \omi / \delta \kxi = \xxi / \tti = v$, but the requirement for the second pulse is $\delta \omi / \delta \kxi = - \xxi /  \tti = -v$.
For a given arbitrary variation, \textit{the global phase introduced cannot vanish for both terms in the distribution}. In other words, when the Gaussian is measured after passing the beam splitter, it can be measured at intervals corresponding to either $+v$ or $-v$, but not both simultaneously.

This does not suggest any mechanism for discontinuous change to the distribution or wavefunction but enforces that what is obtained in measurement must minimize the action. This can be interpreted as an observer-dependent collapse, not for the whole universe but for the participants in the interaction alone. In other words, at each interaction, a particular trajectory is observed and persists for a particular observer.

\subsection{Heisenberg's Uncertainty Principle}

In this section, a brief discussion of the standard uncertainty relations of QM is warranted.

The formulation of quantum mechanics presented here is naturally compatible with Heisenberg's uncertainty relations. The uncertainty relations for Fourier transform pairs are well known.
Generally, an uncertainty principle is true for any dual domains related by the Fourier transform, a result of the Cauchy--Schwartz inequality~\cite{MURTY2019}. Indeed, Heisenberg's seminal paper relied on the Fourier coefficients to demonstrate uncertainty and the need for matrix mechanics~\cite{HEISENBERG1925}. 
Uncertainty relations are an important part of SDT as well because the Fourier transform plays a central role in optical wavefront propagation. 

It is a property of the Fourier transform which leads to the Heisenberg uncertainty relation in quantum mechanics in the first place.
Using the Parseval--Plancherel identity applied to a normalized distribution, Heisenberg showed that~\cite{SITARAM1999} 
\begin{align}
    \label{eqn:FourierUncertainty}
    \lb(\int_{-\infty}^{\infty} dx\, x^2 |f(x)|^2\rb)\lb(\int_{-\infty}^{\infty} dk\, k^2 |\tilde{f}(k)|^2\rb) \geq \frac{1}{(4\pi)^2}.
\end{align} 

Since the momentum operator is associated with the dual of the position operator, the~3+1D distributions in the theory presented here are guaranteed to obey the Heisenberg uncertainty relations during dynamical interactions, since these are enacted via Fourier transforms. 

One careful point needs to be made concerning the unique distinction between coordinates and parameters made in this paper. The notion of ``uncertainty'' corresponds to the \textit{coordinate intervals} of measurement, rather than the parameters. This is apparent from the definition: coordinate intervals are measurable, while parameters are unmeasurable. A related fact is that the integrals in Equation (\ref{eqn:p2Xspace}) are always performed over their entire domain, so there is no way to talk about a "value" or "range of values" of the parameters. Therefore, the discussion of uncertainty refers entirely to coordinate intervals. 
        
\subsection{On Perturbative  Time-Dependent Methods}

The time-dependent Schr\"{o}dinger equation (TDSE) provides a fruitful area of study for this theory, both because it represents a major historical difficulty in our understanding of time and because the adopted solution is based on signal processing.
Here, I will examine the TDSE from a signal perspective and demonstrate its relationship to the proposal at~hand. 

The technique of iteration is used to find the higher order corrections to solutions of the TDSE, as it is for Feynman diagrams or the Dyson series in scattering theory as well. This was done in Section \ref{hdr:twostateSystem} when we examined the equation of motion of the coefficients of spin eigenstates. To obtain more accurate results, one often has to perform many integrals for increasingly diminished returns. In this regard, it is interesting to note in the present theory, instead of time-slicing in order to propagate time, time is part of the solution, just as space is. This approach (\ref{eqn:p2Xspace}) is \textit{not} iterative.
It is an eigenfunction equation to find the 3+1D distributions which are invariant under convolution with a given potential function. This is a benefit of the novel ``outside of'' time approach to describing signals.

If we treat time \textit{as a whole} in this way, instead of iterating gradually over time evolution, it is interesting to wonder whether the higher order perturbative corrections in TDSE, for instance, might converge on a solution that could be found all at once with a single 3+1D convolution. Such a solution might still be solved iteratively, but other possibilities may also be apparent from the new method. We have already seen two examples of this: the energy eigenvalues of the harmonic oscillator were found by insisting that the global phase in the 3+1D distribution vanish, and the resonant mode of a spin system was identified by a cursory look at the Fourier transform of the potential. 

The possibility of providing new ways of solving the TDSE remains speculative.
With this in mind, however, I will briefly review the TDSE and show its similarities with Equation (\ref{eqn:p2Xspace}).
We'll revisit the example of a two-state system with a sinusoidal driving potential. The expression for the coefficients of the eigenstates can be readily seen as a recursion of forward and inverse transforms. For instance, consider the second-order correction, 
\begin{align}
    \bs
    \label{eqn:perturbativeCoefficients}
    c_n^{(2)}(t) &\propto \int_{t_0}^{t} d t_1 \int_{t_0}^{t_1} d t_2 \sum_m\bra{n}e^{i\hat{H}_0t_1}V(t_1)e^{-i\hat{H}_0t_1}\ket{m}
    \bra{m}e^{i\hat{H}_0t_2}V(t_2)e^{-i\hat{H}_0t_2}(e^{i\hat{H}_0t_2}\ket{i}) \\
    &\propto  \int_{t_0}^{t} d t_1 e^{i\om_n t_1}\sum_m e^{-i\om_m t_1} V_{nm}(t_1) \int_{t_0}^{t_1} d t_2 e^{i\om_m t_2} V_{mi}(t_2), 
    \es
\end{align}
where we have rearranged the factors suggestively to show that for each order of perturbation in which we insert a factor of $V_I$, we perform a transform to $\om$-space and then back to $t$-space.

To emphasize the connection to signal processing, we can insert a windowing function in each time integral as we did in Section \ref{hdr:twostateSystem} to turn the time integrals into Fourier transforms,
\begin{align}
    \bs
    \label{}
    c_n^{(2)}&\propto  \int_{-\infty}^{\infty} d t_1 \, rect(\frac{t_1}{t-t_0}-\frac{1}{2})
    e^{i \om_n t_1}\sum_m e^{-i\om_m t_1} V_{nm} \int_{-\infty}^{\infty} d t_2 \, rect (\frac{t_2}{t_1-t_0}-\frac{1}{2})
    e^{i\om_m t_2} V_{mi}.
    \es
\end{align}
The windowing function serves a role similar to that of a low pass filter (but with the time and frequency domains switched) one by one on each instance of the potential, forming a series of factors of convolutions of the form
\begin{align}
    \bs
    \int_{-\infty}^{\infty} d t_r \, rect\lb(\frac{t_r}{t_{r-1}-t_0}-\frac{1}{2}\rb)
    e^{i \om_n t_r}\sum_m e^{-i\om_m t_r} V_{nm}
    \Longrightarrow \tau_- e^{i\om \tau_+/2} sinc(\om \tau_-/2) \ast V_{nm}
    \es
\end{align}
where $\tau_\pm = t_{r-1}\pm t_0$ are defined from the limits of integration for the $r$th time integral.
This structure reflects the mathematics of reconstructing a frequency domain signal $V$ from its samples. 
At every higher order $N$, $\om$-space is thus sampled to a greater resolution than the previous order perturbation.

Thus, perturbative techniques are methods of sampling the potential at high, medium, and low resolution, just as in image processing. They explore the ``resolution'' of the potential function, both in frequency and in time, treating time as a data signal. The strength of the formulation put forward here is that it emphasizes the informational approach and the time-free and space-free descriptions that can provide useful insights.

This discussion highlights the potential usefulness of a framework which treats time (and, by extension, energy) as signals of information that exist \textit{as a whole}.
Characterizing wavefunctions or fields as 3+1D information might encourage us to ask new questions and form new cross-disciplinary collaborations.

Analysis of the proposed method has not been comprehensive.
Other items that could benefit from further analysis include:
\begin{itemize}
    \item Normalization and probabilistic predictions of measurements;
    \item Compatibility of space and time ``coordinate intervals'' with Lorentzian invariant intervals
    in special relativity;
    \item An understanding of mass;
    \item Applying Equation (\ref{eqn:p2Xspace}) to mixed states and density matrices;
    \item Analyzing the compatibility of the Wigner function with the Fourier transform approach.
    \item 
    We have mostly discussed changes to the phase distributions. It has not been examined what sort of dynamics the \textit{magnitude} of the spectrum would correspond to.
\end{itemize}

\section{Summary}
A novel conclusion on the nature of time and spacetime paths as a whole is drawn from non-relativistic quantum wavefunction propagation,  formulated in terms of the mathematics of scalar diffraction theory in $\xma$- and $\km$-space. This approach provides a view of Feynman's path integral formulation without defining paths over time, i.e., without time-slicing. A fundamental distinction is made between parameters and coordinate intervals, both of which play distinct roles in their respective spaces. The approach draws close parallels to the mathematics of holograms. It is an elegant formulation which may provide clarity on aspects of quantum theory that have previously eluded analysis, such as interpretation of the process of quantum collapse and the nature of observable time.

\vspace{6pt} 



\funding{This research was supported by the Foundational Questions Institute (fqxi.org; grant no. FQXi-1801) via funding from the Federico and Elvia Faggin Foundation.}


\acknowledgments{The author is grateful to Jeff Butler, Richard Pham Vo, Ben Savitsky, Jason Elhaderi, Justin Kader, Dominik Safranek, Kai Chung, Daniel Sheehan, Eliahu Cohen, Marcin Nowakowski, Seth Kostek, Federico Faggin, Anthony Aguirre, Jurgen Theiss, Menas Kafatos, Ken Wharton, and Lincoln Carr for helpful comments and feedback. 
}

\conflictsofinterest{The authors declare no conflict of interest.} 

\appendixtitles{yes} 
\appendixstart
\appendix

\section{Equivalence of Quantum Mechanics and Scalar Diffraction Theory~Formulations \label{hdr:equivalence}}

A review of well-known results comparing QWP and SDT will be provided here.

\subsection{Expressing Wavefunction Propagation in Terms of Fourier Transforms \label{hdr:propagatorFree}}

The propagation through space of a wavefunction can be written in terms of Fourier transforms and appropriate phase factors (cf. \cite{ALBEVERIO2008}). 
The updated state of a wavefunction under evolution by an arbitrary unitary operator, $\hat{U}$, is
\begin{align}
    \Psi(x')= \braket{x'|\hat{U}|\Psi} = \int dx \, \braket{x'|\hat{U}|x} \Psi(x),
    \label{eqn:transmission0}
\end{align}
where the propagator or transition amplitude is $\braket{x'|\hat{U}|x}$.

Inserting the Schr\"{o}dinger propagator into Equation (\ref{eqn:transmission0}) in the free particle case, one can write,
\begin{align}
    \bs
    \label{eqn:mainEquation0}
    \Psi(x_2,\delta t) &= \int dx_1 \,e^{-im(x_2-x_1)^2/2\delta t} \Psi(x_1,0)\\
    &= \int d x_1 \lb(\int dk  e^{i k (x_2-x_1)}  e^{i(k^2/2m)\delta t}\rb)\Psi(x_1,0)\\
    &= \int dx_1 \int dk \braket{x_2|k} 
    \braket{k|x_1} e^{i(k^2/2m)\delta t}
    \braket{x_1|\Psi} \\
    &= \int dk \braket{x_2|k} 
    e^{i(k^2/2m)\delta t}
    \int dx_1 \braket{k|x_1} \braket{x_1|\Psi} \\
    &= \scrT^{-1} \Big\{ e^{i(k^2/2m)\delta t}  \scrT \Big\{\Psi(x_1,0)\Big\}\Big\},
    \es
\end{align}
which is in the form of Equation (\ref{eqn:p2Xspace}).
With a potential $\hat{V}$, Equation (\ref{eqn:mainEquation0}) becomes (cf. \cite{SREDNICKI2007} \mbox{(p. 44))}
\vspace{12pt}
\begin{align}
    \bs
    \label{eqn:QFT1}
    \Psi(x_2,\delta t) &= \int dx_1 \braket{x_2|e^{-i\hat{H}\delta t}|x_1}  \braket{x_1|\Psi} \\
    &= \frac{1}{2 \pi} \int dk \int dx_1 \braket{x_2|e^{-i\hat{\nabla}^2 \delta t/(2m)}|k}\braket{k|e^{i\hat{V}\delta t}|x_1}  \braket{x_1|\Psi} \\
    &= \frac{1}{2 \pi} \int dk \, e^{i k x_2} e^{-i(k^2/2m)\delta t} \int d x_1 e^{iV(x_1) \delta t} e^{-i k x_1} \Psi(x_1,0)\\
    &= \scrT^{-1} \Big\{ e^{-i(k^2/2m)\delta t}  \scrT \Big\{e^{iV(x_1) \delta t}\Psi(x_1,0)\Big\}\Big\}.
    \es
\end{align}

Here, we assume that time is continuous so that $\delta t$ is infinitesimal, and $\exp{(i\hat{H}\delta t)}$ can be factored as a consequence of the Baker--Campbell--Hausdorff formula. 
In the above, the momentum operator eigenfunction is $\braket{k|x}=\frac{1}{\sqrt{2 \pi}}\exp{(-i k x)}$, $\hbar$ is set to unity, and $\scrT$ indicates the Fourier transform with normalization factors absorbed into the definition.

Phase factors (whose signs are chosen for convenience) are applied in each domain,
\begin{align}
    \label{eqn:QFTxPhase}
    S_k &= \frac{k^2 \delta t}{2m}\\
    S_x &= V(x) \delta t.
\end{align}

These phase factors carry important information describing the evolution of the~system.

\subsection{Amplitude Transfer and Transmittance Functions in QWP and SDT}

SDT utilizes both $x$-space and $k$-space to describe how a 2D optical wavefront changes as it propagates in a third spatial dimension (an analog of time). Optical wavefront propagation occurs by multiplying its $k$-space representation by an amplitude \textit{transfer} function. 
Thus, the pattern on a screen is given by a convolution (\cite{GOODMAN2004} (p. 67))
\begin{align}
    \bs
    \label{eqn:fresnel}
    g_f(x_i,y_i) &\propto \scrT^{-1}\{\tilde{h}(k_x,k_y)\tilde{g}_t(k_x,k_y)\} \\
                &\propto h \ast g_t
    \es
\end{align}
where the convolution kernel $h(x, y)$ is the impulse response, $\ast$ indicates convolution, tilde indicates the signal in $k$-space, and the convolution theorem was used. $\tilde{h}$ is the amplitude transfer function (or Fourier transformed impulse response) which propagates the wavefront.
Using this technique, any arbitrary filter $\tilde{h}$ can be applied to a signal.

The QWP equivalent of the amplitude transfer function $\tilde{h}$ is the phase factor applied in $k$-space (inside the inverse Fourier transform) in Equations (\ref{eqn:mainEquation0}) and (\ref{eqn:QFT1}). 
In~both instances, propagation of a signal in space involves multiplication by a phase factor \mbox{in~$k$-space}.

Now, we ask the opposite question: what does multiplication by a phase factor in $x$-space represent? In QWP, this involves the potential, as in the last line of Equation (\ref{eqn:QFT1}). 
In SDT, this corresponds to the amplitude \textit{transmittance} function (\cite{GOODMAN2004} (p. 59))
\begin{align}
    \label{eqn:transmittanceFunction}
    g_t(x,y) = g_i(x,y) \ampTransmittance(x,y)
\end{align}
which is used to describe the local wavefront scaling and/or phase shifting by a complex aperture distribution $\ampTransmittance$. In the special case in which the transformation is unitary, such as with a birefringent filter, Equation (\ref{eqn:transmittanceFunction}) is simply multiplication by a phase factor in~$x$-space.

As Goodman notes, due to the convolution theorem, this is equivalent to convolution of the $k$-space signals, $\tilde{g}_t = \tilde{g}_i \ast \tilde{t}_A$, paralleling our manipulations from Equation (\ref{eqn:fresnel}) but in the dual space. Therefore, in SDT, multiplication by a phase factor in $x$-space corresponds to the interaction of an optical wavefront with a screen or aperture.

In comparing Equations (\ref{eqn:QFT1}) and (\ref{eqn:transmittanceFunction}), we make the identification
\begin{align}
    \label{eqn:unitaryTransmittanceFunction}
    \ampTransmittance(x,y) \rightarrow e^{iV(x)\delta t}.
\end{align}

We can therefore relate apertures in SDT to potentials in QWP.

\subsection{Stepwise Comparison between Optical Wavefront Propagation and Quantum Wavefunction Propagation}

One can now use the identifications between SDT and QWP thus far to illustrate that QWP is a process of diffraction. Equation (\ref{eqn:mainEquation0}) for wavefunction propagation of a free particle in QWP will now be derived from the general solution of the diffraction imaging~problem. 

\begin{figure}[H]
  
  \includegraphics[width=250pt]{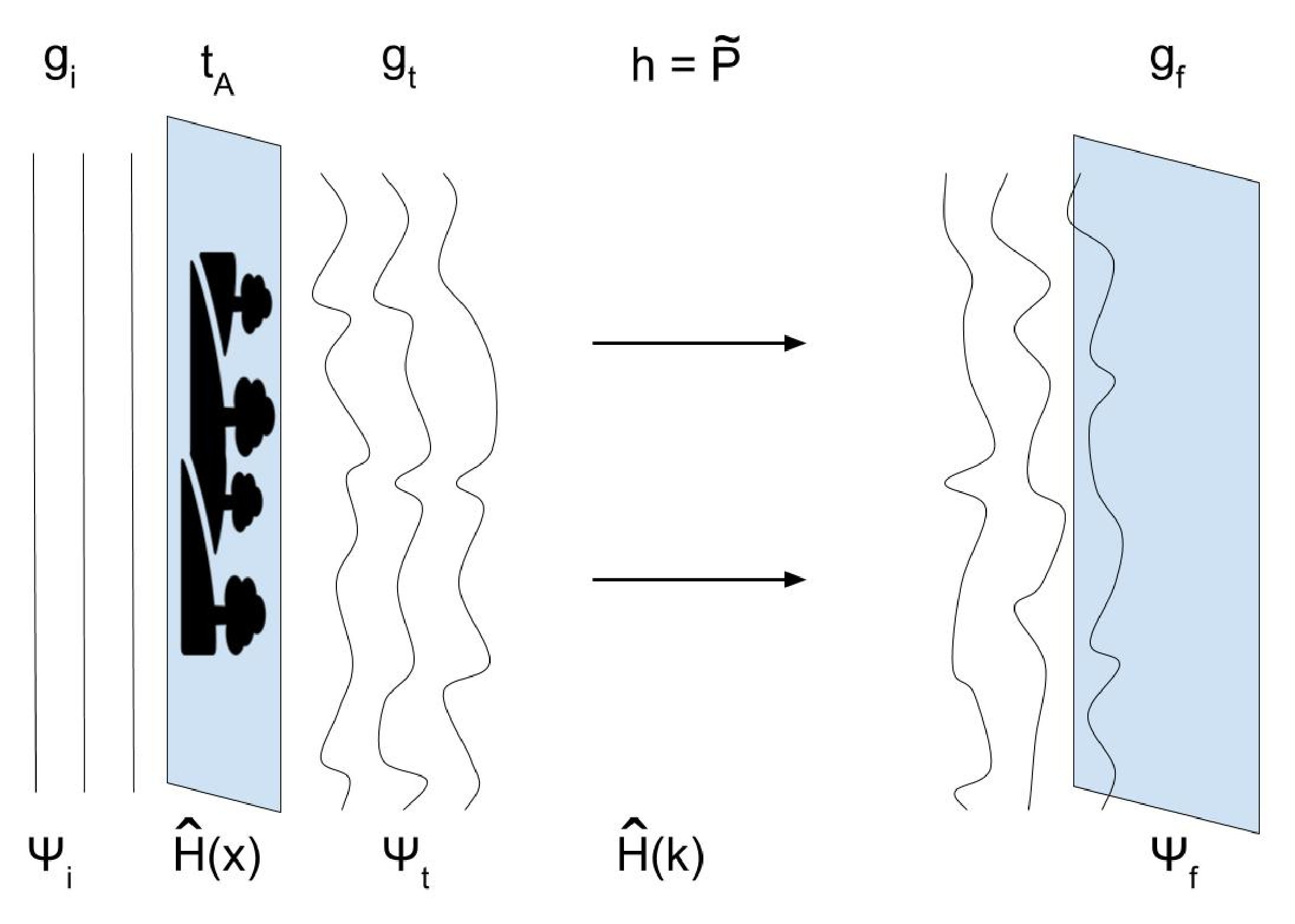}

  \caption{Visual comparison between wavefront propagation in SDT and wavefunction propagation in QM. The incoming wave (SDT optical wavefront, $g_i$, or QM wavefunction, $\Psi_i$) impinges upon a diffractive element (aperture $\ampTransmittance$ to be imaged in SDT, or potential $\hat{H}(x)$ in QWP). The resulting modified waveform ($g_t$ in SDT or $\Psi_t$ in QWP) propagates via application of a phase factor in $k$-space (the impulse response ($h$) or the Fourier-transformed pupil function ($\tilde{P}$) in SDT, or the kinetic term of the Hamiltonian $\hat{H}(k)$ in QWP), resulting in a final waveform ($g_f$ hitting a screen or film in SDT, or~$\Psi_f$ indicating an updated state of the particle in QWP).}
  \label{fig:scalarWavePropagation}
\end{figure}

The propagation of a quantum system through free space can be modeled as an optical wavefront $g_t$ interacting with an imaging system $P$ (e.g., camera). Refer to Figure \ref{fig:scalarWavePropagation}. A signal $g_t$ strikes the camera pupil or aperture, and the physical pupil shape $P(x,y)$ is mapped from $x$-space to frequency space, $P(k_x,k_y)$. The Fourier transform mapping between and $\vec{x}-$space and $\vec{k}-$space parameters is not one-to-one. However, when light passes through an aperture in $\vec{x}-$space, the parameters describing $k$-space are linearly related to those of $x$-space
\begin{align}
    \bs
    \label{eqn:coordinateMapping}
        k_x &= \frac{2\pi x}{\lambda z}\\
        k_y &= \frac{2\pi y}{\lambda z},
    \es
\end{align}
where $z$ is the distance to the screen. 
Because of Equation (\ref{eqn:coordinateMapping}), it is sometimes unclear whether a given expression refers to the frequency space representation or a scaled version of the spatial representation. This is a subtle but important point derived in (\cite{GOODMAN2004} (p. 136)). 
Although it might seem at first glance that the $k$-space representation of a pupil should be the Fourier transform of its shape in $x$-space, the pupil function of a camera actually filters \textit{$k$-space} according to its shape in \textit{$x$-space}. For instance, a small aperture attenuates the high frequencies of a signal because the limited range of the aperture in space gives rise to a correspondingly limited bandwidth. 
The diffractive effect of the pupil function in $x$-space is then characterized by the impulse response, 
\begin{align}
    \label{eqn:pointSpreadFunction}
    h(x,y) = \scrT^{-1} \Big\{P(k_x,k_y) \Big\}.
\end{align}

The final propagated signal is given as a convolution of the impulse response with the original waveform,
\begin{align}
\bs
\label{eqn:transferFunction}
    g_f(x,y) &= h(x,y)\ast g_t(x,y) \\
            &= \scrT^{-1} \Big\{ \scrT \Big\{h(u,v)\Big\} \scrT \Big\{ g_t(u,v)\Big\} \Big\} \\
            &= \scrT^{-1} \Big\{ P(k_x,k_y) \scrT \Big\{ g_t(u,v)\Big\} \Big\},
\es
\end{align}
where the pupil function $P(k_x,k_y)=\tilde{h}(k_x,k_y)$ written in $k$-space, plays the role of the amplitude \textit{transfer} function. 

In SDT, the incoming transverse optical waveform is $g_i(x,y)$, and the wavefront immediately after an aperture is given by $g_t(x,y)$ in Equation (\ref{eqn:transmittanceFunction}). 
Inserting \mbox{Equation (\ref{eqn:transmittanceFunction})} into Equation (\ref{eqn:transferFunction}), relabeling $P$ as $\tilde{h}$, and using the convolution theorem, one obtains
\begin{align}
    \bs
    \label{eqn:SDTPropagationComplete}
    g_f &= \scrT^{-1}\Big\{ \tilde{h}(k_x,k_y) \scrT\Big\{ g_i(x,y)     \ampTransmittance(x,y) \Big\} \Big\}.
    \es
\end{align}

If we choose 
\begin{align}
    \label{eqn:schrodingerTransferFunction}
    \tilde{h}(k_x,k_y) = \exp{\lb(-i\frac{(k_x^2 + k_y^2) \delta t}{2m}\rb)},
\end{align}
\textls[-20]{we have now obtained a form similar to Equation (\ref{eqn:QFT1}), with the identification \mbox{Equation (\ref{eqn:unitaryTransmittanceFunction})}}.
In optics, this transfer function leads to the Fresnel (paraxial/near field) approximation. In~QWP, the equivalent relation is Schr\"{o}dinger's equation. The wavefunction is analogous to the incoming light to be diffracted, the potential function serves as the object causing the diffraction (the diffractive aperture), and the propagation through space is analogous to the pupil, imaging system, or camera.

\section{Comparing the Standard 3D Wavefunction with the 3+1D Wave~Distribution \label{hdr:appB}}

Here will be attempted a cursory comparison between the 3D wavefunction and the 3+1D wave distribution.
The 3D wavefunction in QM has the following properties:

    \begin{Property} It contains all measurable information about a system. \label{itm:enum1}
    \end{Property}
    
    \begin{Property} It is normalizable and can be integrated via Born's rule to obtain the probability of a particular outcome when measuring a system. \label{itm:enum3}
    \end{Property}
    
    \begin{Property} Physical observables are represented by Hermitian operators, and the values that a measurement can yield are given by the eigenvalues associated with each eigenfunction of the~operator. \label{itm:enum2}
    \end{Property}
    
    \begin{Property} The wave function and its derivatives are linear, continuous, and single-valued.\label{itm:enum4}\end{Property}
    
    \begin{Property} As a probability amplitude, it can be used to calculate the expectation value of an observable.\label{itm:enum5}
    \end{Property}
    
    \begin{Property} Through Born's rule, it represents a non-local probability density in space, but it represents a local probability density in time, predicting the probabilities of measurement only at precisely specified times. \label{itm:enum6}
    \end{Property}
    
    \begin{Property} It can be defined separately for each system.\label{itm:enum7}
    \end{Property}
    
    \begin{Property} It evolves via Schr\"{o}dinger's equation.\label{itm:enum8}
    \end{Property}

We now compare these properties to those of the 3+1D wave distributions.

Property \ref{itm:enum1} is postulated to be true by Postulates \ref{pos:p1} and \ref{pos:p2}.

Property \ref{itm:enum3} requires further investigation. 
The 3+1D distributions obey Parseval's theorem, 
\begin{align}
    \int d\xma |\Psi(\xma)|^2 = \int d\km |\tilde{\Psi}(\km)|^2
\end{align}
implying a conservation of probability, similar to a 3D wavefunction. A correspondence in the manner by which one obtains finite probabilities may be possible.

Since there is a single distribution representing the entire $\xma$-space, rather than individual distributions for each subsystem (and similarly for $\km$-space), it is not yet clear whether the normalization condition is required and how it would be interpreted.

To examine Property \ref{itm:enum2} in the new theory, note that the investigation is limited to observables associated with the translation in time, space, energy, and momentum. These Hermitian operators acting on a 3+1D 
ket can be expressed as distributions in the dual spaces using the methods demonstrated in Equation (\ref{eqn:QFT1}). The phase distributions are the eigenvalue distributions of the operators, e.g.,
    \begin{align}
        \bs
        e^{i\hat{p} \Delta x}\ket{\Psi} &= \int dk e^{i\hat{p} \Delta x}\ket{k}\braket{k|\Psi} \\
        &= \int dk e^{ik \Delta x} \tilde{\Psi}\ket{k},
        \es
    \end{align}
where $k$s are the eigenvectors and eigenvalues of $\hat{p}$, $\hat{p}\ket{k}=k\ket{k}$. Thus, the allowable values of a measurement are given by space/time or energy/momentum coordinate pairs that match the path constraints given by Equation (\ref{eqn:p2Xspace}) on these eigenvalues' distributions.

Property \ref{itm:enum4} is true from the postulates, which claim that $\Psi$ and $\tilde{\Psi}$ are distributions.

Property \ref{itm:enum5} needs further exploration than can be accomplished here.

Property \ref{itm:enum6} is a distinction between the 3D case and the 3+1D case. The 3+1D distribution does not have a one-to-one correlation with the geometry of spacetime, just as a holographic plate does not look anything like the image encoded into it. The 3+1D distribution is non-local in both space and time.

\textls[-20]{Property \ref{itm:enum7} is another area of distinction between the 3D and 3+1D cases. \mbox{Equations (\ref{eqn:genericP2})--(\ref{eqn:p2Xspace})}} take in two distributions but put out a single distribution. This works in the 3+1D case because there is a single distribution describing the entire system, rather than separate distributions associated with specific subsystems (although the latter approach can also be used, as it was in Section \ref{hdr:deriveEOM}).

Finally, since Schr\"{o}dinger's equation is a first-order approximation for the second-order time evolution of the wavefunction required by the Klein--Gordon equation, the~paraxial equation may play the same role in SDT for wavefront propagation at small angles as Schr\"{o}dinger's equation plays for small velocities. This is Property \ref{itm:enum8}.

\section{Time as an Observable \label{hdr:AppTime}}

The usage of spacetime eigenvectors $\ket{\xma}$ throughout the paper implies the existence of a time operator for which these are the eigenstates. The difficulties in defining a time operator were discussed early on by Pauli~\cite{PAULI1958} and more recently by Moyer~\cite{MOYER2015}, Pegg~\cite{PEGG1991}, Busch~\cite{BUSCH2008}, and Price~\cite{PRICE1996}. 

My brief treatment will closely follow Moyer, who emphasizes that the difficulties stem from two main issues. 
Firstly, one must ask whether time is an external parameter appearing in Schrodinger's equation or an internal property of the system. Secondly, one~must decide what the crucial function of a time operator would be. Is it conjugate to the Hamiltonian, or is it the generator of time translation?

The first question is addressed in an interesting manner in the present theory. The~parameters $x_i$ and $t_i$, for instance, in Equation (\ref{eqn:dispersionNonRel1}), are continuous eigenvalue distributions from expressions of the form $\partial_x \ket{k_0,\om_0}$ and $\partial_t \ket{k_0,\om_0}$. The time parameter cannot be ``an external parameter according to which the system evolves'', for the 3+1D wave distribution is defined across all of $\xma$-space and cannot itself evolve with time (Corollary \ref{cor:c1}).
Consequently, is the time parameter instead a measurable property of the system? No, the parameters $x_i$ and $t_i$ serve as integration parameters for the integral transform and as such are unmeasurable. 

What about the time \textit{coordinates}, $\xxi$? Are these associated with an external parameter or a property of the system? Coordinate intervals are only defined for interaction events and are therefore not continuous. Measured time coordinate intervals always appear together with measured space coordinate intervals, as in Equation (\ref{eqn:inertialTrajConstraint2}), so while they serve as an external clock, it is wrong to think of the space as dependent on the time in the usual~sense. 

The absence of an external time in the theory presented here is, arguably, a feature rather than a drawback. Pegg argues, ``$\ldots$ the problem (with understanding time as we do space) may arise from the imposition on the system of an external time parameter, with~the Schrodinger equation being applied as a law of evolution in terms of this external time $\ldots$ any reasonable approach should ideally already incorporate Schrodinger’s equation or its equivalent, that is, there should be no need to postulate the form of any time evolution operator, even that applying to the clock''~(\cite{PEGG1991} (p. 1)).

Moyer points out that most efforts to define a time operator start by making it conjugate to a particular Hamiltonian. Instead, to construct time states, he emphasizes the role of the Hamiltonian as the generator of translations in time~\cite{MOYER2015}. He constructs a complete basis of timeline states in Hilbert space which map to the energy eigenstates. He uses these states to establish a ``timeline wave (function)'' worked out explicitly for a free particle in three dimensions. Finally, treating these as eigenstates, he creates a time operator in the usual way and concludes, ``Thus, a canonical time operator for a free particle in three dimensions exists.''

Thus, while the existence of a time operator for the general case is still uncertain, we utilize the eigenvectors $\ket{x,t}$, acknowledging that they are not firmly established in usage.

\section{Interpretations of Quantum Mechanics}

Regarding the measurement and collapse of the wavefunction, this formalism requires a relational or observer-dependent measurement process. It is compatible with any interpretation of QM which views quantum collapse as an observer-dependent process. See \mbox{Table \ref{tab:interpretations}} for a comparison.
        
        \begin{specialtable}[H]
           \caption{Compatibility of interpretations of QM with observer-dependent collapse.} 
           \setlength{\tabcolsep}{18.6mm}
          \begin{tabular}{cc} \toprule
                Many-worlds interpretation&
                compatible\\
                Consistent histories&
                compatible\\
                Relational interpretation&
                compatible\\
                Phase space&
                compatible\\
                Transactional interpretation&
                uncertain\\
                Two-state vector formalism&
                compatible\\
                Penrose interpretation&
                incompatible\\
                GRW&
                incompatible\\
                Copenhagen interpretation&
                agnostic\\
                de Broglie--Bohm theory&
                uncertain\\
               	\bottomrule
            \end{tabular}
            \label{tab:interpretations}
        \end{specialtable}

\subsection{Compatible with Proposed Theory} 
The many-worlds \cite{EVERETT1957} interpretation, also referred to as a relative state theory, is closely linked with the proposal made here. 
Consistent histories makes no assumptions about collapse, simply providing a framework for determining the consistency of chains of events~\cite{HALLIWELL1995,GRIFFITHS1984,GRIFFITHS2002}.
The relational interpretation is premised on the notion of relational collapse, and Rovelli's original paper defines the concept well~\cite{ROVELLI1996,MERMINITHACA1998}. These three well-established approaches to the quantum mechanics formalism are all compatible with the approach put forward here.

The phase space interpretation appears to treat position and momentum space in a manner similar to the approach taken here~\cite{GROENEWOLD1946,MOYAL1949}. The phase space approach in the literature may lead to conclusions that differ somewhat from those emphasized here, such~as that quantum mechanics is at its root a statistical theory. Brown and Hiley developed the Schr\"{o}dinger equation in terms of a phase operator, an approach which seems compatible and worth analyzing in more detail~\cite{BROWN2000}.
Quantum Bayesianism and the Ithaca interpretation~\cite{MERMIN1998} are closely related to the relational approach and highlight correlations (between systems) rather than correlata (the systems in and of themselves) as fundamental. This is compatible with the notion of \textit{coordinates} defined here. 

The transactional interpretation~\cite{CRAMER1986,KASTNER2012} makes the argument that measurements are the results of a complete transaction between start and endpoints, similar to that made here. It is not clear whether the transactional interpretation proposes a single definite and objective measurement outcome for all observers; if so, it would not be compatible. 

The two-state vector formalism \cite{AHARONOV1964} provides a formalism for discussing pre- and post- boundary conditions on histories, looking at the history as a whole. These appear to be compatible with and related to the formalism presented here.

\subsection{Not Compatible with Proposed Theory} 
The Penrose and GRW interpretations explicitly rely on objective collapse as a result of gravity or spontaneous statistics (respectively) and are therefore not compatible with relational collapse. 

The Copenhagen interpretation is not technically an interpretation of quantum mechanics but a framework of mathematical tools. When it is employed in a calculation, a distinction is often made (in the mind of the physicist) between a quantum world and a classical world as distinct regimes. This is supported with the theory of decoherence, which explains macroscopic objectivity under the assumption of a well-defined external environment. This distinction does not exist with relational collapse. In relational models, the environment is also relative and decoherence is an observer-dependent process. Thus, decoherence and relationality are technically compatible, but if the physicist uses decoherence to explain observer-indepedence of measurement, then it becomes incompatible.

The de Broglie--Bohm interpretation appears to claim that a particle is guided by a pilot wave and has a well-defined position at all times, which is counter to the claim here that particles only have definite states at the moments of measurement. However, given~Bohm's views on holograms and what he called the implicate order, more investigation into this interpretation is warranted.


\end{paracol}
\reftitle{References}

\bibliography{main}


\end{document}